\documentclass[a4paper,11pt]{article}
\pdfoutput=1

\usepackage{jheppub}
\usepackage[T1]{fontenc}
\usepackage{amssymb}
\usepackage{graphicx}
\usepackage{amsmath}
\usepackage{hyperref}
\usepackage{tensor}
\usepackage{epstopdf}
\usepackage{extarrows}
\usepackage{verbatim}
\usepackage{subfigure}

\newcommand{\td}{\text{d}}

\begin{document}
	
	\title{Quantization-scheme-Independent Energy and Its Implications for Holographic Bounds}
	
	\author[a]{Ze Li,}
    \author[a] {Hai-Shan Liu,}
    \author[b]{Zi-Qing Xiao,}
    \author[a] {Run-Qiu Yang}
    \emailAdd{lize@tju.edu.cn}
    \emailAdd{hsliu.zju@gmail.com}
    \emailAdd{xiaoziqing24@mails.ucas.ac.cn}
	\emailAdd{aqiu@tju.edu.cn}
	\affiliation[a]{Center for Joint Quantum Studies and Department of Physics, School of Science, Tianjin University, Yaguan Road 135, Jinnan District, 300350 Tianjin, P.~R.~China}
\affiliation[b]{International Centre for Theoretical Physics Asia-Pacific (ICTP-AP), University of Chinese Academy of Sciences (UCAS), 100190 Beijing, China}
	\abstract{In holographic duality, the total energy of the dual field theory is obtained from the holographic renormalization, which depends not only on the bulk geometry but also on the choice of quantization schemes. We point out that the validity of several widely studied holographic inequalities---including the AdS Penrose inequality, the late-time bound on entanglement entropy growth, and the growth-rate limits of CV and CA complexities---depends on the choice of quantization schemes. Motivated by this issue, we introduce a modified total energy, which is still computed via holographic renormalization but the final value is independent of the choice of quantization schemes. We verify that this modified energy removes the apparent violations of these bounds that arise from quantization-scheme dependence in the model of massive scalar field. Our results suggest that our modified total energy provides a more robust notion of energy when we talk about above inequalities in holographic settings.}
	
	% Uncomment for PACS numbers
	%\pacs{00.00, 20.00, 42.10}
	%
	% Uncomment for keywords
	%\vspace{2pc}
	%\noindent{\it Keywords}: XXXXXX, YYYYYYYY, ZZZZZZZZZ
	%
	% Uncomment for Submitted to journal title message
	%\submitto{\JPA}
	%
	% Uncomment if a separate title page is required
	\maketitle
	%
	% For two-column output uncomment the next line and choose [10pt] rather than [12pt] in the \documentclass declaration
	%\ioptwocol
	%
	%\tableofcontents

\section{Introduction}
The AdS/CFT correspondence~\cite{Maldacena:1997re,Gubser1998,Witten1998} provides a powerful framework for understanding strongly coupled field theories via the geometry of asymptotically Anti-de Sitter (AdS) spacetimes. Central to this duality is the holographic dictionary, which relates the asymptotic behavior of bulk fields to data in the boundary conformal field theory (CFT). The precise definition of physical quantities---particularly the total energy (or mass) of the system---is achieved through holographic renormalization~\cite{Balasubramanian:1999re,deHaro2001,Skenderis:2002wp}. By adding appropriate counterterms to the bulk action, one renders the on-shell action finite and obtains a quasilocal boundary stress-energy tensor.

However, a subtle ambiguity arises when considering scalar fields with mass $m$ in the mass window $m^2_{\text{BF}} \leq m^2 < m^2_{\text{BF}} + 1/\ell_{\text{AdS}}^2$, where $m^2_{\text{BF}} = -d^2/(4\ell_{\text{AdS}}^2)$ is the Breitenlohner-Freedman bound~\cite{Breitenlohner:1982jf}, and $d$ is the boundary dimension. Within this range, the scalar field admits two distinct normalizable asymptotic behaviors, characterized by the scaling dimensions $\Delta_+$ (standard quantization) and $\Delta_-$ (alternative quantization)~\cite{Klebanov:1999tb}. The choice between these schemes corresponds to different boundary conditions-Dirichlet versus Neumann/mixed-thereby implying different dual CFTs or double-trace-deformed theories.

Applying holographic renormalization in these two quantization schemes yields different total energies for the same spacetime.
 In the standard quantization (\(\Delta_+\)), the energy is determined by the subleading falloff mode of the scalar field, whereas in the alternative quantization (\(\Delta_-\)) one must introduce an additional boundary term---effectively a Legendre transform---which modifies the on-shell action and shifts the resulting total energy~\cite{Witten2001}. Consequently, even for the same bulk geometry, the total energy extracted holographically depends on the choice of quantization, or equivalently, on the boundary conditions imposed on the scalar field. Given the well-known subtleties surrounding the definition of gravitational energy in general relativity, this scheme dependence is not entirely unexpected.

However, several classical inequalities in general relativity---most notably the Penrose inequality---are highly sensitive to the precise definition of total energy. The Penrose inequality provides a lower bound on the mass of a spacetime in terms of the area \(A\) of its event horizon, $M \geq \left( A/ 16\pi G \right)^{(d-2)/(d-1)}$ as originally proposed in~\cite{Penrose:1973um}. Its AdS generalization takes the form~\cite{Itkin:2011ph}
\begin{equation}
	\label{eq:10001}
	M \geq \left( \frac{A}{16\pi G} \right)^{1/2}
	+ \frac{1}{2\ell^2}\left( \frac{A}{4\pi G} \right)^{3/2},
\end{equation}
and was argued to hold holographically in~\cite{Engelhardt:2019btp}. Differing from the asymptotically flat case, the ``mass'' $M$ in holography is subtle.  In holographic duality, the total energy is usually obtained by holographic renormalization, which is defined as the 00-component of the renormalized Brown--York tensor~\cite{Balasubramanian:1999re,deHaro2001,Skenderis:2002wp}. The right-hand side of inequality~\eqref{eq:10001} is a purely geometric quantity determined by the bulk geometry, while the left-hand side of inequality~\eqref{eq:10001} depends on the choice of quantization schemes, and so leads to a mismatch. Remarkably, Ref.~\cite{Xiao:2022obq} showed that the validity of the AdS Penrose inequality in the presence of scalar fields actually depends on the choice of quantization scheme: the inequality holds under the standard quantization, but apparent violations may arise under the alternative quantization. This mismatch reflects not a failure of the geometric inequality itself, but rather the scheme dependence inherent in the holographic definition of energy.

In addition to the Penrose inequality, the late-time growth rate of holographic entanglement entropy can also admit an upper bound that depends on the total energy. The gravitational dual of the thermofield double (TFD) state is the eternal two-sided AdS black hole obtained from its maximal analytic extension. When computing the holographic entanglement entropy of a TFD state, the corresponding extremal surface may penetrate the black hole horizon~\cite{Hartman2013}, leading to a linear growth at late times. As argued in~\cite{Li:2022cvm}, the growth rate of this entropy is bounded from above; for instance, in planar-symmetric AdS geometries one finds
\begin{equation}
	\label{eq:10002}
	\frac{\mathrm{d}S}{\mathrm{d} t} \leq c_{D}\, E^{1 - \frac{1}{d}},
\end{equation}
where \(c_{D}\) is a constant depending only on the spacetime dimension \(d\).
The numerical analysis in~\cite{Li:2022cvm} shows that, when the energy $E$ is obtained from the usual holographic renormalization scheme, this upper bound is satisfied when the total energy is defined using the standard quantization scheme, whereas it can be violated when the alternative quantization is used.

Another important quantity in holography is the computational complexity. There are two widely studied proposals: the ``complexity = volume'' (CV)~\cite{Susskind:2014rva,Stanford:2014jda} and ``complexity = action'' (CA)~\cite{Brown:2015bva,Brown:2015lvg} holographic complexities. The CV conjecture identifies the complexity of the boundary state with the volume of a codimension-one extremal hypersurface in the bulk, in close analogy with the construction of holographic entanglement entropy. It was further shown in~\cite{Yang:2019alh} that the growth rate of CV complexity admits an upper bound that is proportional to the total energy of the spacetime. In contrast, the CA conjecture associates quantum complexity with the gravitational action evaluated on the Wheeler-DeWitt (WdW) patch~\cite{Brown2016}. The Lloyd bound is often interpreted as an upper bound on the late-time growth rate of holographic complexity:
\begin{equation}
	\label{eq:10004}
	\frac{\td \mathcal{C}}{\td t} \leq \frac{2E}{\pi},
\end{equation}
as argued in~\cite{Lloyd:2000cry}. This upper bound has also been verified in
various holographic setups~\cite{Brown:2015bva,Brown:2015lvg,Lehner:2016vdi,Cai:2016xho,Yang:2016awy}. It is worth noting that the applicability of Lloyd's bound is known to be subtle in more general setups. For example, in charged or rotating black holes, the relevant quantity may involve generalized thermodynamic combinations such as $M-\mu Q-\Omega J$ rather than the total energy alone. In this paper, we examine how the growth-rate bounds of both CV and CA complexities depend on the choice of quantization scheme. As we will show later, if $E$ is obtained by the conventional holographic renormalization scheme, the bound~\eqref{eq:10004} is satisfied under the standard quantization but can be violated under the alternative quantization.

The three inequalities discussed above share several structural features: one side of each inequality depends only on the bulk geometry, while the other side---the total energy---depends on the choice of quantization schemes. When the total energy itself varies with the choice of quantization scheme, the validity of these inequalities becomes scheme-dependent. These facts suggest that the ``total energy'' which should be used in the above inequalities may not be the usual one obtained from holographic renormalization. This naturally leads to the question: should the notion of ``total energy'' appearing in these inequalities be redefined so that they remain valid in a more universal sense?

As a first step, we focus on a simple bulk theory: a free real scalar field in asymptotically AdS spacetime. We introduce a modified total energy $\mathcal{H}$ based on three requirements:
\begin{enumerate}
\item[(1)] in the absence of matter fields, \( \mathcal{H} \) should reduce to the ADM mass;
\item[(2)] \( \mathcal{H} \) should ensure that the above inequalities hold independently of the quantization scheme;
\item[(3)] \( \mathcal{H} \) itself is still obtained from the holographic renormalization but is scheme-independent, i.e.\ it should take the same value regardless of whether one adopts the standard or the alternative quantization.
\end{enumerate}
Based on the above criteria, we propose the following definition of a modified energy:
\begin{equation}
	\label{eq:10003}
	\mathcal{H} = \mathcal{E} + \frac{d - \Delta}{d} \, J \, \langle \mathcal{O} \rangle ,
\end{equation}
where \( \mathcal{E} \) is the holographically renormalized energy that is obtained from the 00-component of holographically renormalized stress-energy tensor, \( J \) is the source, \( \langle \mathcal{O} \rangle \) is the expectation value of the dual operator, and \( \Delta \) denotes its conformal dimension. In Eq.~\eqref{eq:10003}, the $\mathcal{E}$, $J$, $\Delta$ (here $\Delta$ can be either $\Delta_+$ or $\Delta_-$) and $\langle \mathcal{O} \rangle$ are all dependent on the choice of quantization schemes but we will see later that the final value of $\mathcal{H}$ is independent of the quantization schemes. Using this modified energy definition, we demonstrate in this work that all of the inequalities above are restored to being scheme-independent: they hold regardless of which quantization scheme is used.

The remainder of this paper is organized as follows. In Section~\ref{sec002}, we review the standard and alternative quantization schemes and derive the modified holographically renormalized energy. In Section~\ref{sec:example}, we apply this definition to four-dimensional asymptotically AdS spacetime as a concrete example. Section~\ref{sec004} examines the Penrose inequality, the upper bound on the late-time growth rate of holographic entanglement entropy, and the upper bound on the late-time growth rate of holographic complexity. We conclude with a final discussion in Section~\ref{section 8}.

\section{Modified Energy $\mathcal{H}$}\label{sec002}
In the context of AdS/CFT, we consider a $(d+1)$-dimensional bulk theory. As a preliminary study, we focus on a bulk theory consisting of gravity minimally coupled to a free massive scalar field. The \textit{bulk} action $S$ for gravity minimally coupled to a massive scalar field in AdS space is given by
\begin{equation}
	\label{eq:1001}
	S=\frac{1}{16\pi G}\int \mathrm{d}^{d+1} x \sqrt{-g}\left(R-2\Lambda-\frac{1}{2} \nabla^{\mu} \phi \nabla_{\mu} \phi-\frac{1}{2}m^2\phi^2\right),
\end{equation}
where $\Lambda=-\frac{d(d-1)}{2\ell_{\text{AdS}}^{2}}$ denotes the cosmological constant, and $m$ is the mass of the scalar field $\phi$.
Varying the action with respect to $\phi$, we obtain the equation of motion:
\begin{equation}
	\label{eq:1002}
	\nabla_{\mu} \nabla^{\mu} \phi-m^2\phi=0.
\end{equation}
Using the separation of variables for the scalar field, we focus on its radial part $\phi(r)$. We assume that the dual boundary theory is in a flat spacetime. Due to the asymptotically AdS nature of the spacetime, the leading-order bulk metric is assumed to take the following form near the AdS boundary
\begin{equation}\label{adsmetric}
  \mathrm{d}s^2=\frac{r^2}{\ell_{\text{AdS}}^{2}}\mathrm{d}s^2_{\mathrm{flat}}+\frac{\ell_{\text{AdS}}^{2}\mathrm{d}r^2}{r^2}\,.
\end{equation}
Here $\mathrm{d}s^2_{\mathrm{flat}}$ is the $d$-dimensional Minkowski metric. The scalar field equation near the boundary ($r \to \infty$) reduces to
\begin{equation}
	\label{eq:1003}
	\frac{1}{r^{d-1}}\partial_{r}(r^{d+1}\partial_{r}\phi(r))-m^2 \ell_{\text{AdS}}^{2} \phi(r)=0.
\end{equation}
The solution can be expanded near the AdS boundary ($r \rightarrow \infty$) as
\begin{equation}
	\label{eq:1004}
	\phi(r,\vec{x})=\frac{\phi_{\alpha}}{r^{\Delta_{-}}}\left(1+\cdots\right)+\frac{\phi_{\beta}}{r^{\Delta_{+}}}\left(1+\cdots\right),
\end{equation}
where $\phi_{\alpha}(\vec{x})$ and $\phi_{\beta}(\vec{x})$ are functions of the boundary coordinates $\vec{x}$. The exponents $\Delta_{\pm}$ are determined by the characteristic equation $\Delta(\Delta-d) = m^2\ell_{\text{AdS}}^{2}$, whose solutions are
\begin{equation}
	\label{eq:1005}
	\Delta_{\pm}=\frac{d}{2} \pm \frac{1}{2}\sqrt{d^2+4m^2\ell_{\text{AdS}}^{2}}.
\end{equation}

According to the AdS/CFT dictionary, the exponents $\Delta_{\pm}$ are related to the conformal dimensions of operators in the boundary conformal field theory. For the mass of the scalar field in the range $m^2_{\text{BF}} \leq m^2 < m^2_{\text{BF}} + 1/\ell_{\text{AdS}}^2$, where $m^2_{\text{BF}} = -d^2/(4\ell_{\text{AdS}}^2)$ is the Breitenlohner-Freedman bound, and provided the corresponding conformal dimensions $\Delta_{\pm}$ satisfy the unitarity bound ($\Delta \geq \frac{d-2}{2}$), both branches of the solution are theoretically allowed. These two cases correspond to the standard quantization scheme (where the boundary operator has dimension $\Delta_+$) and the alternative quantization scheme (where the boundary operator has dimension $\Delta_-$), respectively.

In the bulk, these two quantization schemes correspond to different boundary conditions for the scalar field at the asymptotic boundary. The action~\eqref{eq:1001} only includes the bulk terms necessary to derive the equations of motion. To have a well-defined variational principle and to render the on-shell action finite, one must add boundary and counterterms. This process is known as holographic renormalization. The full renormalized action is given by
\begin{equation}
	\label{eq:1006}
	\begin{aligned}
		&S_{\text{tot}} = S_{\text{Bulk}} + S_{\text{GHY}} + S_{\text{ct, gravity}} + S_{\text{ct, scalar}}, \\
		&S_{\text{Bulk}} = \frac{1}{16\pi G}\int_{\mathcal{M}} \mathrm{d}^{d+1} x \sqrt{-g}\left(R-2\Lambda-\frac{1}{2} \nabla^{\mu} \phi \nabla_{\mu} \phi-\frac{1}{2}m^2\phi^2\right), \\
		&S_{\text{GHY}} = \frac{1}{8\pi G}\int_{\partial \mathcal{M}} \mathrm{d}^{d} x \varepsilon\sqrt{|h|} K, \\
		&S_{\text{ct, gravity}} =- \frac{1}{8\pi G}\int_{\partial \mathcal{M}} \mathrm{d}^{d} x \varepsilon\sqrt{|h|} K_{0}, \\
		&S_{\text{ct, scalar}} = -\frac{1}{16\pi G}\int_{\partial \mathcal{M}} \mathrm{d}^{d} x \sqrt{|h|} \frac{\Delta_{-}}{2 \ell_{\text{AdS}}}\phi^2.
	\end{aligned}
\end{equation}
Here, $h$ is the induced metric on the boundary $\partial \mathcal{M}$, $K$ is the trace of the extrinsic curvature, $\varepsilon$ is 1 for timelike segments of $\partial\mathcal{M}$ and $-1$ for spacelike segments of $\partial\mathcal{M}$, and $K_0$ is a gravitational counterterm designed to cancel the divergent contributions from the bulk and GHY terms. $S_{\text{ct, scalar}}$ is the counterterm for the scalar field. The special cases in which $\Delta_{+}-\Delta_{-}=2\nu$ with $\nu\in \mathbb{Z}_{+}$ lead to logarithmic subleading terms and necessitate extra counterterms in the renormalization procedure~\cite{Balasubramanian:1999re,deHaro2001,Skenderis:2002wp}. In the present paper we avoid these complications and consider only the non-resonant regime $\Delta_{+}-\Delta_{-}\neq 2\nu$. \footnote{Here we used the minimal subtraction for the action of scalar field. }
Throughout this paper, $\mathcal{M}$ denotes an arbitrary $(d+1)$-dimensional subregion of AdS, with $\partial \mathcal{M}$ its boundary. When $\partial \mathcal{M}$ is the $r\to\infty$ hypersurface, $\mathcal{M}$ becomes the full AdS spacetime.
Now consider the variation of the total action with respect to the scalar field $\phi$. The contribution from the bulk and scalar counterterm is:
\begin{equation}
	\label{eq:1007}
	\begin{aligned}
		\delta S_{\text{tot}}(\delta \phi) &= \delta S_{\text{Bulk}} + \delta S_{\text{ct, scalar}} \\
		&=\frac{1}{16\pi G}\int_{\mathcal{M}} \mathrm{d}^{d+1} x \sqrt{-g} (\text{E.o.M.})\delta \phi -\frac{1}{16\pi G}\int_{\partial \mathcal{M}} \mathrm{d}^{d} x \sqrt{|h|} \left( n^{\mu} \nabla_{\mu} \phi + \frac{\Delta_{-}}{ \ell_{\text{AdS}}}\phi \right) \delta \phi.
	\end{aligned}
\end{equation}
Here ``E.o.M.'' stands for the equations of motion, and $n^{\mu}$ denotes the outward-pointing unit normal vector to the boundary $\partial\mathcal{M}$. Imposing Dirichlet boundary conditions amounts to fixing the value of $\phi$ on $\partial\mathcal{M}$, so that $\delta\phi|_{\partial \mathcal{M}}=0$. Together with the on-shell condition $\text{E.o.M.} = 0$, the total variation $\delta S_{\text{tot}}$ vanishes, ensuring a well-defined variational principle.
Since the bulk contribution vanishes once the equations of motion are imposed, the on--shell action reduces to a functional depending only on boundary data, such as $\phi_{\text{boundary}}$ and $h_{\mu\nu}$. On the boundary $\partial \mathcal{M}$ we may furthermore introduce the momentum conjugate to $\phi_{\text{boundary}}$, without considering the holographic renormalization, defined by~\cite{Papadimitriou:2004ap}
\begin{equation}
	\label{eq:1008}
	\Pi = -\frac{16\pi G}{\sqrt{-h}}\frac{\delta S_{\text{Bulk, on-shell}}}{\delta \phi_{\text{boundary}}} = n^{\mu} \nabla_{\mu} \phi\,.
\end{equation}
In the holographic framework, one typically works with the renormalised action $S_{\text{tot}}$. Correspondingly, we define the renormalised boundary conjugate momentum $\Pi_{\text{ren}}$ by including the contribution from the scalar counterterm:
\begin{equation}
	\label{eq:1009}
	\Pi_{\text{ren}} = -\frac{16\pi G}{\sqrt{-h}}\frac{\delta S_{\text{tot, on-shell}}}{\delta \phi_{\text{boundary}}} = n^{\mu} \nabla_{\mu} \phi + \frac{\Delta_{-}}{ \ell_{\text{AdS}}}\phi.
\end{equation}
Substituting the asymptotic expansion \eqref{eq:1004} into the expression for $\Pi_{\text{ren}}$, we find that its leading-order behavior is
\begin{equation}
	\label{eq:1010}
	\Pi_{\text{ren}} = \frac{1}{ \ell_{\text{AdS}}}\frac{(\Delta_{-}-\Delta_{+})\phi_{\beta}}{ r^{\Delta_{+}}} + \cdots.
\end{equation}
This expression turns out to be particularly useful for subsequent computations near the asymptotic AdS boundary.

At the boundary $\partial \mathcal{M}$, in addition to the Dirichlet boundary condition ($\delta \phi = 0$), one can also impose a Neumann boundary condition by fixing the renormalized momentum ($\delta \Pi_{\text{ren}} = 0$). However, under the Neumann boundary condition, the boundary term $-\Pi_{\text{ren}} \delta \phi$ in the variation of the action does not vanish. Consequently, the variational principle is not well-posed for fixed $\Pi_{\text{ren}}$ since $\delta S_{\text{tot}} \neq 0$ even when the equations of motion hold.
To obtain a well-posed variational principle with fixed $\Pi_{\text{ren}}$, we can add a boundary term that does not alter the equations of motion, known as the Legendre term~\cite{Klebanov:1999tb}:
\begin{equation}
	\label{eq:1011}
	S_{\text{Legendre}} = \frac{1}{16\pi G} \int_{\partial \mathcal{M}} \mathrm{d}^{d} x \sqrt{-h}  \Pi_{\text{ren}}  \phi.
\end{equation}
The variation of the new total action, $S_{\text{tot}}^{\text{(alt)}} = S_{\text{tot}} + S_{\text{Legendre}}$, is then given by
\begin{equation}
	\label{eq:1012}
	\delta S_{\text{tot}}^{\text{(alt)}} = \delta S_{\text{Bulk}} + \delta S_{\text{ct, scalar}} + \delta S_{\text{Legendre}} = \frac{1}{16\pi G}\int_{\partial \mathcal{M}} \mathrm{d}^{d} x \sqrt{-h}  \phi  \delta \Pi_{\text{ren}},
\end{equation}
here we have used the on-shell condition $\text{E.o.M.} = 0$. It is then evident from Eq.~\eqref{eq:1012} that, upon imposing the Neumann boundary condition ($\delta \Pi_{\text{ren}} = 0$), the variation of the action $S_{\text{tot}}^{\text{(alt)}}$ vanishes and the variational principle is well defined.

Let us now specialize to the case in which $\partial M$ is the asymptotic AdS boundary at $r\to\infty$. From Eqs.~\eqref{eq:1004} and~\eqref{eq:1010} one could conclude that both $\phi$ and its momentum $\Pi$ vanish in this limit, which would make the choice of boundary conditions rather subtle. Interestingly, substituting the asymptotic expansions~\eqref{eq:1004} and~\eqref{eq:1010} into Eqs.~\eqref{eq:1007} and~\eqref{eq:1012} shows that the variation of the action takes the schematic form
\begin{equation}
	\label{eq:100121}
	\begin{aligned}
		\delta S_{\text{tot}} & \propto \phi_{\beta} \delta \phi_{\alpha}, \\
		\delta S_{\text{tot}}^{\text{(alt)}} & \propto \phi_{\alpha} \delta \phi_{\beta}.
	\end{aligned}
\end{equation}
 In other words, the usual Dirichlet condition $\delta\phi=0$ is effectively replaced by $\delta\phi_{\alpha}=0$, while the Neumann condition $\delta\Pi_{\text{ren}}=0$ corresponds instead to $\delta\phi_{\beta}=0$. This makes it clear that the appropriate boundary conditions at infinity should be formulated in terms of the coefficients $\phi_{\alpha}$ and $\phi_{\beta}$ rather than $\phi$ and $\Pi_{\mathrm{ren}}$.

We now identify the boundary CFT interpretation.
In the standard quantization scheme, one imposes the Dirichlet boundary condition by fixing $\phi_{\alpha}$ at the boundary. In the dual CFT, $\phi_{\alpha}$ corresponds to the source $J_\alpha$ coupled to a boundary operator $\mathcal{O}_{\alpha}$ of dimension $\Delta_+$, and the expectation value $\langle \mathcal{O}_{\alpha} \rangle$ is related to $\phi_{\beta}$.
Conversely, in the alternative quantization scheme, one imposes the Neumann boundary condition by fixing $\Pi_{\text{ren}}$. This is equivalent to fixing $\phi_{\beta}$, which now serves as the source $J_\beta$ for a boundary operator $\mathcal{O}_{\beta}$ of dimension $\Delta_-$, with its expectation value $\langle \mathcal{O}_{\beta} \rangle$ related to $\phi_{\alpha}$.
Thus, from the bulk perspective, the two quantization schemes are related by a Legendre transformation, implemented by the boundary term $S_{\text{Legendre}}$. This fundamental difference will be reflected in the definitions of the renormalized boundary stress-energy tensor and the expectation values of the dual scalar operators $\mathcal{O}_{\alpha}$ and $\mathcal{O}_{\beta}$.

For the standard quantization scheme, the holographic stress-energy tensor is defined as
\begin{equation}
	\label{eq:1013}
	T_{ij}^{\text{(sta)}} = \frac{2}{\sqrt{-h}}\frac{\delta S_{\text{tot}}^{\text{(sta)}}}{\delta h^{ij}}
	= \frac{2}{\sqrt{-h}}\frac{\delta (S_{\text{GHY}} + S_{\text{ct, gravity}} + S_{\text{ct, scalar}})}{\delta h^{ij}}.
\end{equation}
The expectation value of the dual scalar operator $\mathcal{O}_{\alpha}$ is given by
\begin{equation}
	\label{eq:1014}
	\langle \mathcal{O}_{\alpha} \rangle = \frac{1}{\sqrt{-h}}\frac{\delta S_{\text{tot}}^{\text{(sta)}}}{\delta J_{\alpha}}
	= \frac{1}{\sqrt{-h}}\frac{\delta S_{\text{tot}}^{\text{(sta)}}}{\delta \phi_{\alpha}} = (\Delta_{+} - \Delta_{-}) \phi_{\beta}.
\end{equation}
For the alternative quantization scheme, the holographic stress-energy tensor is
\begin{equation}
	\label{eq:1015}
	T_{ij}^{\text{(alt)}} = \frac{2}{\sqrt{-h}}\frac{\delta S_{\text{tot}}^{\text{(alt)}}}{\delta h^{ij}}
	= \frac{2}{\sqrt{-h}}\frac{\delta (S_{\text{GHY}} + S_{\text{ct, gravity}} + S_{\text{ct, scalar}} + S_{\text{Legendre}})}{\delta h^{ij}}.
\end{equation}
The expectation value of the dual scalar operator $\mathcal{O}_{\beta}$ is given by
\begin{equation}
	\label{eq:1016}
	\langle \mathcal{O}_{\beta} \rangle = \frac{1}{\sqrt{-h}}\frac{\delta S_{\text{tot}}^{\text{(alt)}}}{\delta J_{\beta}}
	= \frac{1}{\sqrt{-h}}\frac{\delta S_{\text{tot}}^{\text{(alt)}}}{\delta \phi_{\beta}} = -(\Delta_{+} - \Delta_{-}) \phi_{\alpha}.
\end{equation}
From above expressions on dual boundary stress-energy tensor, it is not surprising that the same bulk geometry will lead to two different values of dual boundary energy in two different quantization schemes.

However, recent investigations into various holographic quantities-such as the Penrose inequality, holographic entanglement entropy, and complexity (under both the CV and CA conjectures)-have related the quantities defined by bulk geometries to the ``energy'' of dual boundary theory according to a few of these inequalities. Although these inequalities are supported to hold in holography, the meaning of ``energy'' in these inequalities is subtle. In holography, we should use the holographic renormalization to obtain the boundary stress-energy tensor. However, if we then naively use the 00-component to obtain the ``energy'' and put it into the above inequalities, we then find a significant dependence on the choice of boundary quantization scheme. Studies~\cite{Xiao:2022obq, Li:2022cvm, Yang:2019alh} demonstrate that the definitions of the stress-energy tensor in the standard and alternative quantization schemes lead to vastly different physical results in these contexts.
This inconsistency motivates us to suspect: the ``energy'' in above inequalities is not directly obtained from the 00-component of renormalized stress-energy tensor. Instead, we may seek a definition of a new modified energy functional, $\mathcal{H}$, whose value is invariant under the choice of quantization scheme, thereby providing a unified description for various physical situations.

We start from the variational expression for the conserved charge at infinity, \(\delta H_{\infty}\), constructed using the Wald formalism. As shown in Ref.~\cite{Li:2020spf}, its variation satisfies
\begin{equation}
	\label{eq:101701}
	\frac{\delta H_{\infty}}{\Sigma}
	= \delta \mathcal{E} + \langle \mathcal{O} \rangle \, \delta J .
\end{equation}
Here \( \mathcal{E} \) denotes the energy extracted from the \(00\)-component of the holographically renormalized stress-energy tensor
\(T_{ij}\) given in \eqref{eq:1013} or \eqref{eq:1015}, with \(\mathcal{E} = 16\pi T_{00}\).
The quantity \( J \) represents the source, identified as either \(J=\phi_{\alpha}\) or \(J=\phi_{\beta}\), while \(\langle \mathcal{O} \rangle\) is the expectation value of the dual scalar operator defined in \eqref{eq:1014} or \eqref{eq:1016}. Here $\mathcal{E}$, $\langle \mathcal{O} \rangle$ and  $J$ will all depend on the choice of quantization schemes. $\Sigma$ denotes the transverse volume. One can verify that the form $\delta H_{\infty}$ is identical for both standard and alternative quantization schemes. However, this form is not closed ($\delta^2 H_{\infty} \neq 0$), which prevents its direct integration to define a conserved charge $H_{\infty}$ that can be identified with the total energy.
A natural approach is to add terms to $\delta H_{\infty}$ to obtain a closed form. The most intuitive modification is to subtract $\langle \mathcal{O} \rangle \delta J$, which yields
\begin{equation}
	\label{eq:1018}
	\frac{\delta \tilde{H}}{\Sigma} = \frac{\delta H_{\infty}}{\Sigma} - \langle \mathcal{O} \rangle \delta J = \delta \mathcal{E}.
\end{equation}
This form is closed and integrates to $\mathcal{E}$, which is nothing but the 00-component of renormalized stress-energy tensor. However, while this yields an energy $\mathcal{E}$, its value itself depends on the choice of quantization scheme, which is the inconsistency we seek to resolve.
We note that the choice of terms to add in order to achieve a closed form is not unique. Here, we propose a different modification that symmetrically incorporates data from both quantization schemes. We add the term $-\left(\frac{\Delta_{+}}{d}\langle \mathcal{O}_{\alpha} \rangle \delta J_{\alpha} + \frac{\Delta_{-}}{d}\langle \mathcal{O}_{\beta} \rangle \delta J_{\beta}\right)$. The resulting closed form is
\begin{equation}
	\label{eq:1019}
	\delta \mathcal{H} = \delta \mathcal{E} + \langle \mathcal{O} \rangle \delta J -\left(\frac{\Delta_{+}}{d}\langle \mathcal{O}_{\alpha} \rangle \delta J_{\alpha} + \frac{\Delta_{-}}{d}\langle \mathcal{O}_{\beta} \rangle \delta J_{\beta}\right).
\end{equation}
We will in the following prove that (1) it is a closed form, and (2) it is invariant under the choice of quantization scheme. Thus, the integration of this closed form defines the modified energy $\mathcal{H}$, which is invariant under the choice of quantization scheme.

For the standard and alternative quantization schemes, the variation $\delta \mathcal{H}$ is given by
\begin{equation}
	\label{eq:1020}
	\begin{aligned}
		\delta \mathcal{H}^{\text{(sta)}} &= \delta \mathcal{E}^{\text{(sta)}} + \langle \mathcal{O}_{\alpha} \rangle \delta J_{\alpha} -\left(\frac{\Delta_{+}}{d}\langle \mathcal{O}_{\alpha} \rangle \delta J_{\alpha} + \frac{\Delta_{-}}{d}\langle \mathcal{O}_{\beta} \rangle \delta J_{\beta}\right)\\
		&= \delta \mathcal{E}^{\text{(sta)}} + \frac{d-\Delta_{+}}{d}\left(\langle \mathcal{O}_{\alpha} \rangle \delta J_{\alpha} -\langle \mathcal{O}_{\beta} \rangle \delta J_{\beta}\right),\\
		\delta \mathcal{H}^{\text{(alt)}} &= \delta \mathcal{E}^{\text{(alt)}} + \langle \mathcal{O}_{\beta} \rangle \delta J_{\beta} -\left(\frac{\Delta_{+}}{d}\langle \mathcal{O}_{\alpha} \rangle \delta J_{\alpha} + \frac{\Delta_{-}}{d}\langle \mathcal{O}_{\beta} \rangle \delta J_{\beta}\right)\\
		&= \delta \mathcal{E}^{\text{(alt)}} + \frac{d-\Delta_{-}}{d}\left(\langle \mathcal{O}_{\alpha} \rangle \delta J_{\alpha} -\langle \mathcal{O}_{\beta} \rangle \delta J_{\beta}\right).
	\end{aligned}
\end{equation}
Recalling the identifications $J_{\alpha} = \phi_{\alpha}$, $\langle \mathcal{O}_{\alpha} \rangle = (\Delta_{+}-\Delta_{-})\phi_{\beta}$, $J_{\beta} = \phi_{\beta}$, and $\langle \mathcal{O}_{\beta} \rangle = -(\Delta_{+}-\Delta_{-})\phi_{\alpha}$, we find that the combination $\langle \mathcal{O}_{\alpha} \rangle \delta J_{\alpha} - \langle \mathcal{O}_{\beta} \rangle \delta J_{\beta}$ is a total variation:
\begin{equation}
	\label{eq:10201}
	\langle \mathcal{O}_{\alpha} \rangle \delta J_{\alpha} - \langle \mathcal{O}_{\beta} \rangle \delta J_{\beta} = (\Delta_{+} - \Delta_{-}) \delta(\phi_{\alpha}\phi_{\beta}) = \delta(J\langle \mathcal{O} \rangle),
\end{equation}
where $J\langle \mathcal{O} \rangle=J_{\alpha}\langle \mathcal{O}_\alpha \rangle$ or $J\langle \mathcal{O} \rangle =J_\beta\langle \mathcal{O}_\beta \rangle$ is a scheme-independent bilinear form. This key property ensures that $\delta \mathcal{H}$ is a closed form. Now Eq.~\eqref{eq:1020} becomes
\begin{equation}\label{eq:1020b}
  \delta \mathcal{H}^{\text{(sta)}}=\delta \left[\mathcal{E}^{\text{(sta)}} + \frac{d-\Delta_{+}}{d}J\langle \mathcal{O} \rangle\right],~\delta \mathcal{H}^{\text{(alt)}}=\delta \left[\mathcal{E}^{\text{(alt)}} + \frac{d-\Delta_{-}}{d}J\langle \mathcal{O} \rangle\right]\,.
\end{equation}
Integrating $\delta \mathcal{H}$ then yields the modified energy in a unified form:
\begin{equation}
	\label{eq:1021}
	\mathcal{H} = \mathcal{E} + \frac{d - \Delta}{d} J \langle \mathcal{O} \rangle.
\end{equation}
Note that here $\mathcal{E}, \Delta, \langle \mathcal{O} \rangle$ and $J$ all depend on the choice of quantization schemes.

We now verify the consistency of the modified energy $\mathcal{H}$ by explicitly checking its invariance under the choice of quantization scheme. The difference in the original energy $\mathcal{E}$ between the two schemes originates from the Legendre term $S_{\text{Legendre}}$ in the total action. This difference is given by
\begin{equation}
	\label{eq:1022}
	\mathcal{E}^{\text{(alt)}} - \mathcal{E}^{\text{(sta)}} = \frac{1}{16 \pi G \Sigma} \frac{2}{\sqrt{-h}}\frac{\delta S_{\text{Legendre}}}{\delta h^{tt}} = (\Delta_{+} - \Delta_{-})  \phi_{\alpha} \phi_{\beta}.
\end{equation}
Next, we compute the difference in the additional term $\frac{d-\Delta}{d} J \langle \mathcal{O} \rangle$ between the two schemes:
\begin{equation}
	\label{eq:1023}
	\left( \frac{d-\Delta_{-}}{d} J_{\beta} \langle \mathcal{O}_{\beta} \rangle \right) - \left( \frac{d-\Delta_{+}}{d} J_{\alpha} \langle \mathcal{O}_{\alpha} \rangle \right) = - (\Delta_{+} - \Delta_{-})  \phi_{\alpha} \phi_{\beta}.
\end{equation}
Adding the results from Eqs.~\eqref{eq:1022} and~\eqref{eq:1023}, we find the total difference in the modified energy:
\begin{equation}
	\mathcal{H}^{\text{(alt)}} - \mathcal{H}^{\text{(sta)}} = \left( \mathcal{E}^{\text{(alt)}} - \mathcal{E}^{\text{(sta)}} \right) + \left[ \left( \frac{d-\Delta_{-}}{d} J_{\beta} \langle \mathcal{O}_{\beta} \rangle \right) - \left( \frac{d-\Delta_{+}}{d} J_{\alpha} \langle \mathcal{O}_{\alpha} \rangle \right) \right] = 0.
\end{equation}
This confirms that $\mathcal{H}$ is indeed identical in both quantization schemes, as required.
The above construction defines $\mathcal{H}$ purely from boundary data. In Appendix~\ref{appendix A} we show that $\mathcal{H}$ can also be understood as the boundary limit of a radially conserved Noether charge of emerging scaling symmetry in the bulk, which provides a complementary symmetry interpretation.

\section{The four-dimensional AdS spacetime with a massive scalar field}\label{sec:example}
Having established the formalism of two holographic renormalization schemes and defined the modified energy $\mathcal{H}$ according to Eq.~\eqref{eq:1021} in the previous section, we now apply these concepts to a concrete example: a four-dimensional AdS spacetime coupled to a massive scalar field. We consider the action~\eqref{eq:1001} with $d=3$, $G=1$ and $\ell_{\text{AdS}} = 1$ for simplicity. To ensure that both standard and alternative quantization schemes are admissible, we choose a scalar mass within the range $-9/4 < m^2 < -5/4$. As a specific case satisfying this bound, we set $m^2 = -2$ throughout this section. Varying the action with respect to the metric $g_{\mu\nu}$ and the scalar field $\phi$ yields the equations of motion:
\begin{equation}
	\label{eq:3001}
	\begin{aligned}
		R_{\mu \nu} - \frac{1}{2} R g_{\mu \nu} + \Lambda g_{\mu \nu} &= \frac{1}{2} \left( \nabla_{\mu} \phi \nabla_{\nu} \phi - \frac{1}{2} g_{\mu \nu} \nabla_{\rho} \phi \nabla^{\rho} \phi - \frac{1}{2} g_{\mu \nu} m^2 \phi^2 \right), \\
		\nabla_{\mu} \nabla^{\mu} \phi - m^2 \phi &= 0.
	\end{aligned}
\end{equation}
We adopt the following planar symmetric ansatz:
\begin{equation}
	\label{eq:3002}
	\begin{aligned}
		\td s^{2} &= -f(r) e^{-\chi(r)} \td t^{2} + \frac{ \td r^{2}}{f(r)} + r^{2} \td \mathbf{x}_{d-1}^2, \\
		\phi &= \phi(r).
	\end{aligned}
\end{equation}
Here, $f$, $\chi$, and $\phi$ are functions of the radial coordinate $r$ only, and $\td \mathbf{x}_{d-1}^2 = \td x_{1}^{2} + \td x_{2}^{2} + \cdots + \td x_{d-1}^{2}$ is the line element of a $(d-1)$-dimensional planar subspace. With this ansatz, the equations of motion \eqref{eq:3001} reduce to the following system of ordinary differential equations:
\begin{equation}
	\label{eq:3003}
	\begin{aligned}
		\frac{\chi^{\prime}}{r} + \frac{1}{d-1} \phi^{\prime 2} &= 0,\\
		\frac{2}{r} \frac{f^{\prime}}{f} - \frac{\chi^{\prime}}{r} - \frac{2 d}{f} + \frac{1}{d-1} \frac{m^2\phi^2}{f} + \frac{2(d-2)}{r^{2}} &= 0, \\
		\frac{f^{\prime \prime}}{f} - \chi^{\prime \prime} + \frac{1}{2} \chi^{\prime 2} + \frac{(d-2)\chi^{\prime}}{r} + \left(\frac{d-3}{r} - \frac{3}{2} \chi^{\prime}\right) \frac{f^{\prime}}{f} - \frac{2(d-2)}{r^{2}} &= 0, \\
		\phi^{\prime \prime} + \left( \frac{f^{\prime}}{f} - \frac{\chi^{\prime}}{2} + \frac{d-1}{r} \right) \phi^{\prime} - \frac{m^2\phi}{f} &= 0.
	\end{aligned}
\end{equation}
Since there are only three independent functions ($f$, $\chi$, $\phi$), only three of the equations in \eqref{eq:3003} are independent.\footnote{One can verify that the fourth equation can be derived from the first three.} For our subsequent numerical calculations, it is convenient to work with the first, second, and fourth equations. The third equation will be automatically satisfied by solutions to this chosen set and can serve as a consistency check.

We require the solution to be asymptotically AdS. This imposes the following boundary conditions on the metric functions and scalar field as $r \to \infty$:
\begin{equation}
	\label{eq:3004}
	\lim_{r \to \infty} f(r) = \frac{r^2}{\ell_{\text{AdS}}^{2}}, \qquad
	\lim_{r \to \infty} \chi(r) = 0, \qquad
	\lim_{r \to \infty} \phi(r) = 0.
\end{equation}
Near the AdS boundary ($r \to \infty$), the scalar field $\phi(r)$ admits an asymptotic expansion of the form
\begin{equation}
	\label{eq:3005}
	\phi(r) = \frac{\phi_{\alpha}}{r^{\Delta_{-}}} \left(1 + \cdots \right) + \frac{\phi_{\beta}}{r^{\Delta_{+}}} \left(1 + \cdots \right),
\end{equation}
where $\phi_{\alpha}$ and $\phi_{\beta}$ are constants parameterizing the boundary data. The scaling dimensions $\Delta_{\pm}$, given by Eq.~\eqref{eq:1005}, evaluate to
\begin{equation}
	\label{eq:3006}
	\Delta_{-} = 1, \qquad \Delta_{+} = 2.
\end{equation}

In this framework, black hole solutions exist provided there is a horizon at some finite radius $r = r_h > 0$, defined by $f(r_h) = 0$. The temperature $T$ and entropy density $\mathcal{S}$ of the black hole are then given by
\begin{equation}
	\label{eq:3007}
	T = \frac{1}{4 \pi} f^{\prime}(r_h) \mathrm{e}^{-\chi(r_h)/2}, \qquad \mathcal{S} = \frac{1}{4} r_h^{d-1}.
\end{equation}

Solving the equations of motion \eqref{eq:3003} perturbatively near the AdS boundary yields the following asymptotic expansions for $f(r)$, $\chi(r)$, and $\phi(r)$~\cite{Li:2020spf}:
\begin{equation}
	\label{eq:3008}
	\begin{aligned}
		\phi(r) &= \frac{\phi_{\alpha}}{r} + \frac{\phi_{\beta}}{r^{2}} - \frac{\phi_{\alpha}^3}{8 r^{3}} + \mathcal{O}(1/r^{4}), \\
		f(r) &= r^2 \left[ 1 + \frac{\phi_{\alpha}^2}{4 r^{2}} + \frac{f_3}{r^{3}} + \mathcal{O}(1/r^{4}) \right], \\
		\chi(r) &= \frac{\phi_{\alpha}^2}{4 r^{2}} + \frac{2\phi_{\alpha}\phi_{\beta}}{3 r^{3}} + \mathcal{O}(1/r^{4}).
	\end{aligned}
\end{equation}
The expansion contains three independent parameters: $\phi_{\alpha}$, $\phi_{\beta}$, and $f_3$. These parameters ultimately determine the physical properties of the solution, such as the energy (mass), black hole temperature and horizon radius of the spacetime. In the following, we will compute the quantities $\mathcal{E}$, $\langle \mathcal{O} \rangle$, and the modified energy $\mathcal{H}$ according to different quantization schemes, explicitly verifying scheme-dependence of $\mathcal{H}$.

In the standard quantization scheme, the source is identified as $J=J_{\alpha} = \phi_{\alpha}$. The holographic stress-energy tensor, given by Eq.~\eqref{eq:1013}, evaluates to
\begin{equation}
	\label{eq:3009}
	T_{ij}^{\text{(sta)}} = \frac{1}{16\pi} \lim_{r \to \infty} r \left[ 2\left(K h_{i j} - K_{i j} - 2 h_{i j}\right) - \frac{1}{2} h_{i j} \phi^{2} \right].
\end{equation}
For a static spacetime, the energy density is given by the $T_{00}$ component of the stress-energy tensor. Thus, we find
\begin{equation}
	\label{eq:3010}
	\mathcal{E}^{\text{(sta)}} = 16 \pi T_{00}^{\text{(sta)}} = -2 f_3 + \phi_{\alpha} \phi_{\beta}.
\end{equation}
The one-point function of the boundary operator $\mathcal{O}_{\alpha}$, according to Eq.~\eqref{eq:1014}, is
\begin{equation}
	\label{eq:3011}
	\langle \mathcal{O}_{\alpha} \rangle = \phi_{\beta}.
\end{equation}
Substituting these results into the definition of the modified energy in Eq.~\eqref{eq:1021} ($\Delta = \Delta_+ = 2$, $d=3$), we obtain
\begin{equation}
	\label{eq:3012}
	\mathcal{H}^{\text{(sta)}} = \mathcal{E}^{\text{(sta)}} + \frac{d - \Delta_{+}}{d} J_{\alpha} \langle \mathcal{O}_{\alpha} \rangle = -2 f_3 + \frac{4}{3} \phi_{\alpha} \phi_{\beta}.
\end{equation}

In the alternative quantization scheme, the source is identified as $J=J_{\beta} = \phi_{\beta}$. The holographic stress-energy tensor, given by Eq.~\eqref{eq:1015}, evaluates to
\begin{equation}
	\label{eq:3013}
	T_{ij}^{\text{(alt)}} = \frac{1}{16\pi} \lim_{r \to \infty} r \left[ 2\left(K h_{ij} - K_{ij} - 2 h_{ij}\right) + h_{ij} \left( \phi n^{a} \partial_{a}\phi + \frac{1}{2} \phi^{2} \right) \right].
\end{equation}
The corresponding energy density is then given by
\begin{equation}
	\label{eq:3014}
	\mathcal{E}^{\text{(alt)}} = 16 \pi T_{00}^{\text{(alt)}} = -2 f_3 + 2\phi_{\alpha} \phi_{\beta}.
\end{equation}
We explicitly see from Eqs.~\eqref{eq:3010} and \eqref{eq:3014} that the renormalized energy $\mathcal{E}$ is different in two quantization schemes when $\phi_\alpha\phi_\beta\neq0$. The one-point function of the boundary operator $\mathcal{O}_{\beta}$, according to Eq.~\eqref{eq:1016}, is
\begin{equation}
	\label{eq:3015}
	\langle \mathcal{O}_{\beta} \rangle = -\phi_{\alpha}.
\end{equation}
The modified energy in this scheme is therefore
\begin{equation}
	\label{eq:3016}
	\mathcal{H}^{\text{(alt)}} = \mathcal{E}^{\text{(alt)}} + \frac{d - \Delta_{-}}{d} J_{\beta} \langle \mathcal{O}_{\beta} \rangle = -2 f_3 + \frac{4}{3} \phi_{\alpha} \phi_{\beta}.
\end{equation}
This result confirms the invariance of the modified energy: $\mathcal{H}^{\text{(sta)}} = \mathcal{H}^{\text{(alt)}}$.

In this section, we have examined a specific model of four-dimensional AdS spacetime coupled to a massive scalar field. Through explicit calculation, we have verified that the modified energy $\mathcal{H}$ is indeed identical in both the standard and alternative quantization schemes.
In the following section, we will investigate the role of $\mathcal{H}$ in several key holographic probes: the Penrose inequality, holographic entanglement entropy, and holographic complexity. Our goal is to determine whether employing $\mathcal{H}$ resolves the quantization-scheme dependence of these quantities reported in earlier studies~\cite{Xiao:2022obq, Li:2022cvm, Yang:2019alh}.

\section{The Application of Modified Energy in Different Problems}\label{sec004}

\subsection{Penrose Inequality}\label{sec-Penrose-ineq}

The weak cosmic censorship conjecture (WCCC) posits that singularities formed through gravitational collapse are always concealed behind an event horizon, preventing distant observers from accessing them~\cite{Penrose:1969pc}. As an extension of the WCCC, Roger Penrose proposed that the total mass of a spacetime and the area of its event horizon should satisfy an inequality, now known as the Penrose inequality~\cite{Penrose:1973um}:
\begin{equation}
	M \geq \sqrt{\frac{A}{16\pi}},
\end{equation}
with equality holding if and only if the spacetime is that of a Schwarzschild black hole. This inequality implies that, for a given event-horizon area, the total mass of the system is bounded from below: one cannot ``support'' an arbitrarily large event horizon with an arbitrarily small mass. It reflects the intrinsic relationship between mass and spacetime curvature: producing stronger gravitational effects (sufficient to form a black hole of a given size) requires at least a corresponding amount of energy. Though the rigorous proof of this inequality is still absent, under certain mathematical and physical assumptions, proofs of this inequality have been established~\cite{Bray:1999eow,huisken2001inverse}. A few recent processes can be found in Refs.~\cite{Itkin:2011ph,Engelhardt:2019btp,Lee:2015xha,Husain:2017cmj,McCormick:2019fie,Mai:2020sac} and references therein.

When the cosmological constant is negative ($\Lambda < 0$), the spacetime structure changes fundamentally. For an asymptotically Schwarzschild-AdS spacetime with curvature radius $\ell$ (where $\Lambda = -3/\ell^2$), the inequality must include an additional term. The conjectured Penrose inequality in AdS generalizes to~\cite{Itkin:2011ph}:
\begin{equation}
	M \geq \left( \frac{A}{16\pi} \right)^{1/2} + \frac{1}{2 \ell^2} \left( \frac{A}{4\pi} \right)^{3/2} .
\end{equation}
The saturation of this inequality corresponds to the Schwarzschild-AdS black hole solution. This bound reflects the interplay between gravitational attraction and the confining potential of the AdS background.

In holography, the bulk's asymptotically black hole brane is dual to the QFT states on the two asymptotic boundaries~\cite{Witten:1998qj,Aharony:1999ti}. The entropy of the reduced density matrix for each boundary is proportional to the area of the black hole horizon~\cite{Ryu:2006bv,Engelhardt:2017aux}. Assuming a given total mass $M$ and the holographic principle, Ref.~\cite{Marolf:2018ldl} shows that  boundaries' QFT state dual to Schwarzschild-AdS black hole has the maximum entropy. Consequently, the AdS Penrose inequality can be derived from the above basic holographic argument:
\begin{equation}\label{cge}
A \le \max A = A_\text{sch}.
\end{equation}
Here $A_\text{sch}$ stands for the horizon area of Schwarzschild-AdS black hole with same total energy. This idea was recently used by Ref.~\cite{Engelhardt:2019btp} to argue the Penrose inequality in asymptotically
AdS spacetime.
However, to make sense in general AAdS spacetimes, a robust definition of total energy is required to verify the Penrose inequality. In AAdS spacetimes, the standard ADM formulation is often insufficient due to divergences at the asymptotic boundary. The procedure of holographic renormalization provides a systematic remedy~\cite{Skenderis:2002wp}. The total energy is given by the $T_{00}$ component of the holographic stress-energy tensor, as shown in Equations~\eqref{eq:3010} and~\eqref{eq:3011}. When the mass of the matter field satisfies $m^2_{\text{BF}} \leq m^2 < m^2_{\text{BF}} + 1/\ell_{\text{AdS}}^2$, two distinct quantization schemes exist, resulting in two different total energies. This begs a question: when we talk about the Penrose inequality in asymptotically, which quantization scheme does the term ``mass'' in the sentence ``black hole with a given mass'' refer to?

Taking the four-dimensional spacetime adopted in Section~\ref{sec:example} as an example, our total energy density is expressed as
\begin{equation}
	\label{eq:7003}
	\frac{\tilde{f}_{3}}{2} = 4 \pi T_{00} =
	\begin{cases}
		- \frac{f_3}{2} + \frac{\phi_{\alpha} \phi_{\beta}}{4}, & \text{standard quantization}\\[2mm]
		- \frac{f_3}{2} + \frac{\phi_{\alpha} \phi_{\beta}}{2}, & \text{alternative quantization}
	\end{cases}.
\end{equation}
Reference~\cite{Xiao:2022obq} shows that, Penrose inequality is satisfied under the standard quantization scheme, but it clearly fails under the alternative quantization scheme. This brings us into a strange situation: the two quantization schemes are both allowed in holography, but only one can satisfy the Penrose inequality. However, the arguments in Refs.~\cite{Marolf:2018ldl,Engelhardt:2019btp} does not rely on the choice of quantization and so implies Penrose inequality should hold in both quantization schemes.

\begin{figure}[h]
	\centering
	\includegraphics[width=0.7\textwidth]{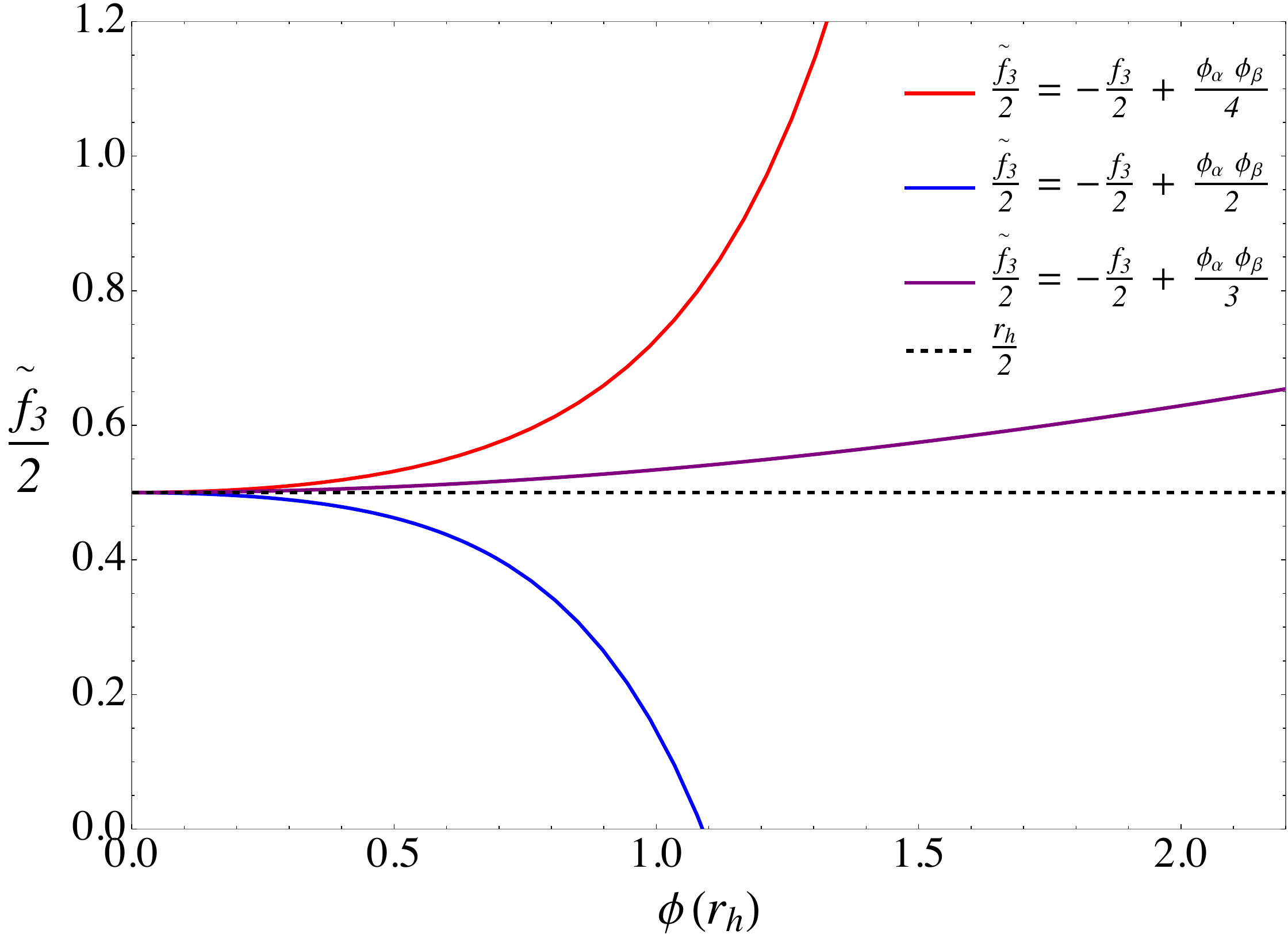}
	\caption{The total energy parameter $\tilde{f}_{3}/2$ as a function of $\phi(r_{h})$. The red and blue curves correspond to the mass density obtained by standard and alternative quantization schemes, respectively. The purple curve represents the case where the modified energy $\mathcal{H}$ is used. The dashed line represents $r_{h}/2$.}
	\label{Fig4004}
\end{figure}

This is where our defined modified energy, $\mathcal{H}$, comes into play. Using this modified energy, Equation~\eqref{eq:7003} is replaced by
\begin{equation}
	\label{eq:7005}
	\frac{\tilde{f}_{3}}{2} = \frac{\mathcal{H}}{4} = - \frac{f_3}{2} + \frac{\phi_{\alpha} \phi_{\beta}}{3},
\end{equation}
which is independent of the quantization scheme. If we use this modified energy as the ``mass'' in the Penrose inequality, then we should expect the following inequality for model in Sec.~\ref{sec:example}
\begin{equation}\label{eq:7004}
  \frac{\tilde{f}_{3}}{2}\geq \frac{r_h}{2}\,.
\end{equation}
Here $\tilde{f}_{3}$ is given according to Eq.~\eqref{eq:7005}. In order to check above inequality, we performed numerical calculations and plotted the total energy parameter $\tilde{f}_{3}/2$ against the scalar field strength $\phi(r_{h})$, as shown in Figure~\ref{Fig4004}. The horizon radius was fixed at $r_{h} = 1$, so the dashed line represents $r_{h}/2 = 1/2$, while the red and blue curves correspond to the standard and alternative quantization schemes, respectively. As noted in Ref.~\cite{Xiao:2022obq}, the former satisfies the Penrose inequality, whereas the latter does not. However, once we adopt our newly defined modified energy $\mathcal{H}$, which is independent of the quantization scheme, inequality~\eqref{eq:7004} is satisfied, as expected.

\subsection{Growth Rate of Holographic Entanglement Entropy}\label{sec:HEE_growth}
Our second example is the entropy growth rate of Thermofield Double (TFD) state in holography. Entanglement entropy serves as a fundamental measure of quantum correlations in many-body systems. Within the AdS/CFT correspondence, this quantity acquires a profound geometric interpretation known as the Ryu-Takayanagi (RT) formula~\cite{Ryu:2006bv}. Consider a CFT residing on the asymptotic boundary of an AdS spacetime. If we define a subregion $A$ on the boundary, the entanglement entropy $S_A$ of this region is given by the area of a minimal surface $\gamma_A$ in the bulk:
\begin{equation}
	S_{\text{HEE}} = \frac{\text{Area}(\gamma_A)}{4G_N},
\end{equation}
where $G_N$ is the Newton constant in the bulk gravity theory. The surface $\gamma_A$ is defined as the codimension-2 minimal surface that satisfies the homology constraint: it must share the same boundary as $A$ ($\partial \gamma_A = \partial A$) and be homologous to $A$. For time-dependent spacetimes, the formula is covariantized into the Hubeny-Rangamani-Takayanagi (HRT) formula~\cite{Hubeny:2007xt}, where the ``minimal'' surface is replaced by an ``extremal'' surface in the Lorentzian geometry.

A particularly rich setting for studying holographic entanglement is the eternal black hole in AdS. The maximally analytically continued Schwarzschild-AdS geometry, which contains two asymptotic boundaries, is dual to TFD state of two entangled CFTs:
\begin{equation}
	|\text{TFD}\rangle = \frac{1}{\sqrt{Z}} \sum_n e^{-\beta E_n / 2} |n\rangle_L \otimes |n\rangle_R.
\end{equation}
We consider the time-evolution of this TFD state $|\text{TFD}(t)\rangle$ under the Hamiltonian of $H=H_L+H_R$, where $H_L$ and $H_R$ stand for the Hamiltonian of left boundary and the right boundary fields. We then trace out one side and compute the entanglement entropy $S$ of this reduced density matrix. In holography, this time-dependent entanglement entropy is dual to an extremal surface $\gamma$ that is enclosed by the two equal-time slices of two boundaries. In this setup, the extremal surface $\gamma$ traverses the bulk, connecting the two asymptotic boundaries through the black hole interior.

As the area of the extremal surface evolves over time due to its passage through the black hole interior~\cite{Hartman2013}, the growth rate of entanglement entropy is given by
\begin{equation}
	\label{eq:40021}
	\frac{\text{d} S}{\text{d} t} \propto G(r_{A}) = \sqrt{-f(r_{A}) \, \text{e}^{-\chi(r_{A})}} \, r_{A}^{\,d-1},
\end{equation}
where $r_A$ is the minimal radial coordinate of extremal surface inside the event horizon. At late times, it saturates to the maximal value $G_{\text{max}} = G(r_{A,m})$, where $r_{A,m}$ is the value that maximizes the function $G(r_{A})$.

Reference~\cite{Li:2022cvm} investigates the upper bound of this growth rate in holographic theories. It first considers the asymptotically Schwarzschild-AdS spacetime, where functions $f(r)$ and $\chi(r)$ in metric~\eqref{eq:3002} at AdS boundary $r\rightarrow\infty$ satisfy:
	\begin{equation}
	\label{eq:40021b}
		f(r)=\frac{r^2}{\ell_{\rm AdS}^{2}}-f_{0}/r^{d-2}+\mathcal{O}(1/r^{d-1}), \qquad
		\chi(r)=\mathcal{O}(1/r^{d+1}).
	\end{equation}
Here $f_0$ just gives the total mass (density) of spacetime. It demonstrates that, for a system with a fixed total energy, there exists an upper bound on the entanglement growth rate. Analysis shows that Schwarzschild-AdS black holes satisfy this constraint in all possible states for a given energy density. Specifically, for a planar-symmetric black hole (corresponding to a flat boundary theory in $d$ dimensions), the maximum growth rate of entanglement entropy is given by
\begin{equation}
	\label{eq:40506}
	\frac{\td S}{\td t} \leq
	\frac{\Sigma_{d-2}}{2 G_{N}} \sqrt{\frac{d}{d-2}} \left(\frac{2(d-1)}{d-2}\right)^{\frac{1}{d}-1} \ell_{\rm AdS}^{1-\frac{2}{d}} f_{0}^{1-\frac{1}{d}},
\end{equation}
where $f_{0}$ is the total mass (energy) density of the spacetime, $d$ is the spacetime dimension of the boundary CFT, and $\Sigma_{d-2}$ is the transverse volume.

However, the ``asymptotically Schwarzschild-AdS condition''~\eqref{eq:40021b} is too strong in holography and is not satisfied in most situations. When  matter does not decay rapidly enough, an asymptotically AdS spacetime may not be asymptotically Schwarzschild-AdS. For example, Eq.~\eqref{eq:3008} does not satisfy the asymptotically Schwarzschild-AdS condition when scalar field is nonzero. Reference~\cite{Li:2022cvm} also considers a more complex scenario in which the AdS black hole spacetime includes a real scalar field. In this context, the definition of total energy depends on the choice of boundary conditions, resulting in two distinct quantization schemes. Under the standard quantization scheme, the conclusion remains consistent with the pure gravity case: the Schwarzschild-AdS geometry yields the maximum growth rate. However, under the alternative quantization scheme, the opposite occurs: the Schwarzschild-AdS geometry exhibits the \textit{minimum} growth rate.

In this section, we investigate whether the above mismatch could be resolved when the total energy is defined by the modified energy $\mathcal{H}$, which is invariant under the choice of quantization scheme. We work within the same model setup as in Sec.~\ref{sec:example}. Our objective is to compute the relationship between the maximum growth rate $G_{\text{max}}$ and $\mathcal{H}$ for different scalar field configurations parameterized by $\phi_{\alpha}$ and $\phi_{\beta}$.

\begin{figure}[h]
	\centering
	\includegraphics[width=0.7\textwidth]{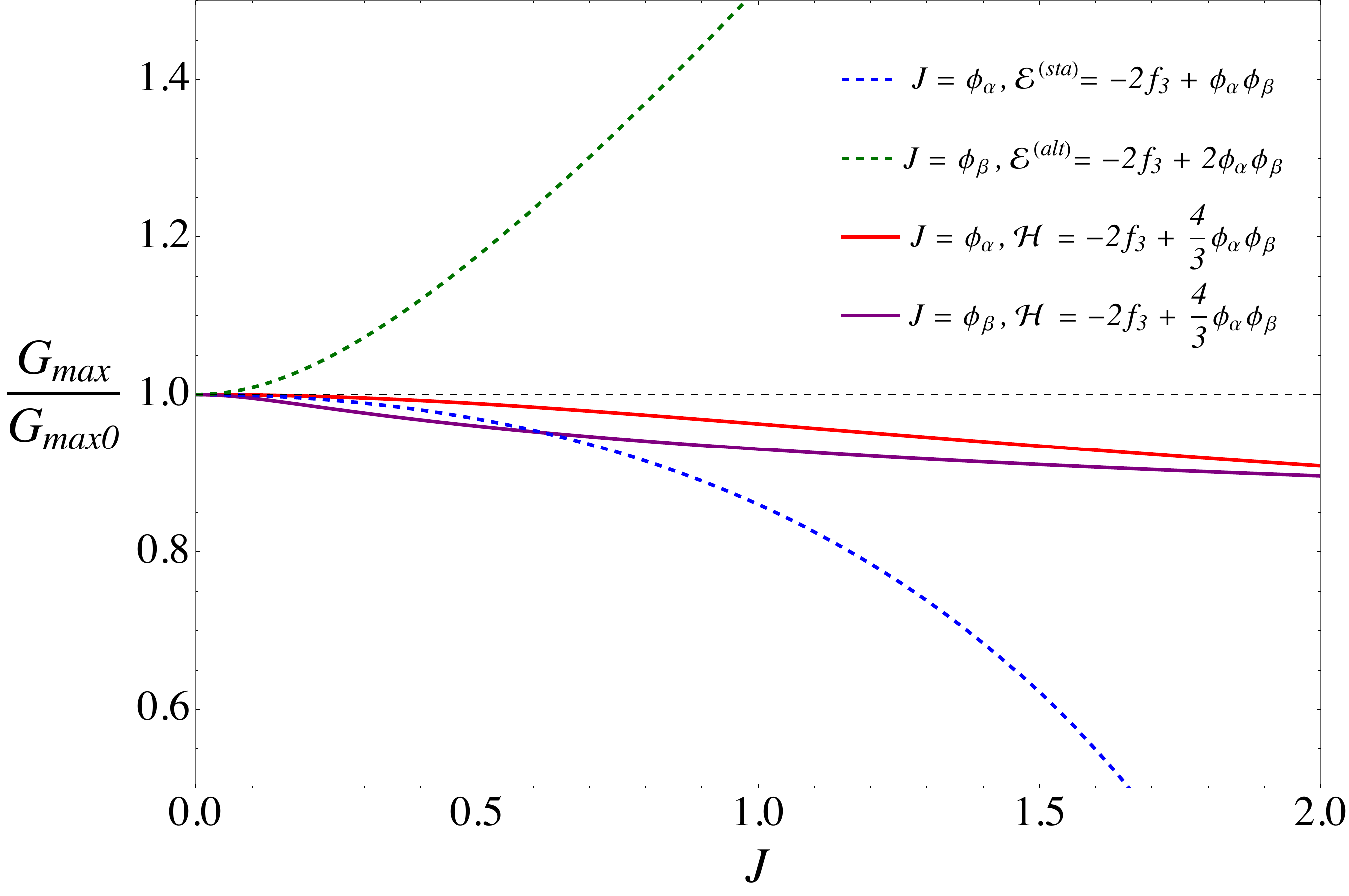}
	\caption{Late-time growth rate of holographic entanglement entropy $G_{\text{max}}$ as a function of the source $J$ with fixed energies under different quantization schemes. The vertical axis shows the normalized growth rate $G_{\text{max}} / G_{\text{max}0}$, where $G_{\text{max}0}$ is the value for the vacuum black hole. The blue and green dashed curves correspond to the standard and alternative quantization schemes with fixed $\mathcal{E}^{\text{(sta)}}$ and $\mathcal{E}^{\text{(alt)}}$, respectively. The red and purple solid curves represent cases with fixed modified energy $\mathcal{H}$.}
	\label{Fig4001}
\end{figure}

As shown in Fig.~\ref{Fig4001}, the vertical axis, $G_{\text{max}} / G_{\text{max}0}$, represents the normalized growth rate, where $G_{\text{max}0}$ corresponds to the vacuum black hole (i.e., with vanishing scalar field). When the modified energy $\mathcal{H}$ is held fixed, $G_{\text{max}}$ attains its maximum value if and only if the scalar field vanishes. This confirms that the maximality principle is restored when the total energy is defined by the modified energy $\mathcal{H}$.

\subsection{Growth Rate of Complexity in CV Conjecture}\label{sec:CV_complexity}
The third example that we will check is the complexity growth rate in the CV conjecture. The CV conjecture of complexity~\cite{Susskind:2014rva,Stanford:2014jda} posits a direct geometric dual to the computational complexity of a boundary quantum field theory state. It states that the complexity $\mathcal{C}$ of a state on a spacelike boundary slice $\Sigma$ is proportional to the volume of the maximal codimension-one hypersurface $\mathcal{B}$ anchored on that slice and extending into the bulk:
\begin{equation}
	\mathcal{C}_V(\Sigma) = \max_{\partial \mathcal{B} = \Sigma} \left[ \frac{V(\mathcal{B})}{G_N \ell} \right],
\end{equation}
where $V(\mathcal{B})$ is the volume of the bulk hypersurface $\mathcal{B}$, $G_N$ is Newton's constant, and $\ell$ is a length scale associated with the bulk geometry, such as the horizon radius or AdS radius. The maximization over all such hypersurfaces ensures the selection of the extremal (maximal) volume slice.

For an eternal AdS black hole dual to a TFD state, the extremal hypersurface $\mathcal{B}$ is analogous to the extremal surface $\gamma$ used for holographic entanglement entropy in the previous subsection, except that $\mathcal{B}$ is codimension-1 while $\gamma$ is codimension-2. The late-time growth rate function is similarly defined as in~\eqref{eq:40021}:
\begin{equation}
	\label{eq:40031}
	\frac{\text{d} \mathcal{C}_{V}}{\text{d} t} \propto \sigma(r_{A}) = \sqrt{-f(r_{A}) \, \text{e}^{-\chi(r_{A})}} \, r_{A}^{d}.
\end{equation}
To be a well-defined complexity, it is expected to satisfied the Lloyd's bound $\text{d} \mathcal{C}/\text{d} t \leq 2E/\pi$. Reference~\cite{Yang:2019alh} demonstrates that this is true for static planar symmetric asymptotically Schwarzschild-AdS black hole under dominant energy condition. However, as with entanglement entropy, the presence of a massive scalar field breaks the asymptotically Schwarzschild-AdS condition and leads to a scheme-dependent definition of the total energy $\mathcal{E}$. Consequently, the Lloyd's bound holds only for the standard quantization scheme, while an apparent violation emerges under alternative quantization
due to the scheme dependence of the holographic energy.

In this section, we investigate whether the Lloyd's bound for complexity growth is restored when using the modified energy $\mathcal{H}$. We employ the same model and parameters as in Sec.~\ref{sec:example} and analyze the relationship between $\sigma_{\text{max}}$ and $\mathcal{H}$ for various scalar field configurations.
\begin{figure}[h]
	\centering
	\includegraphics[width=0.7\textwidth]{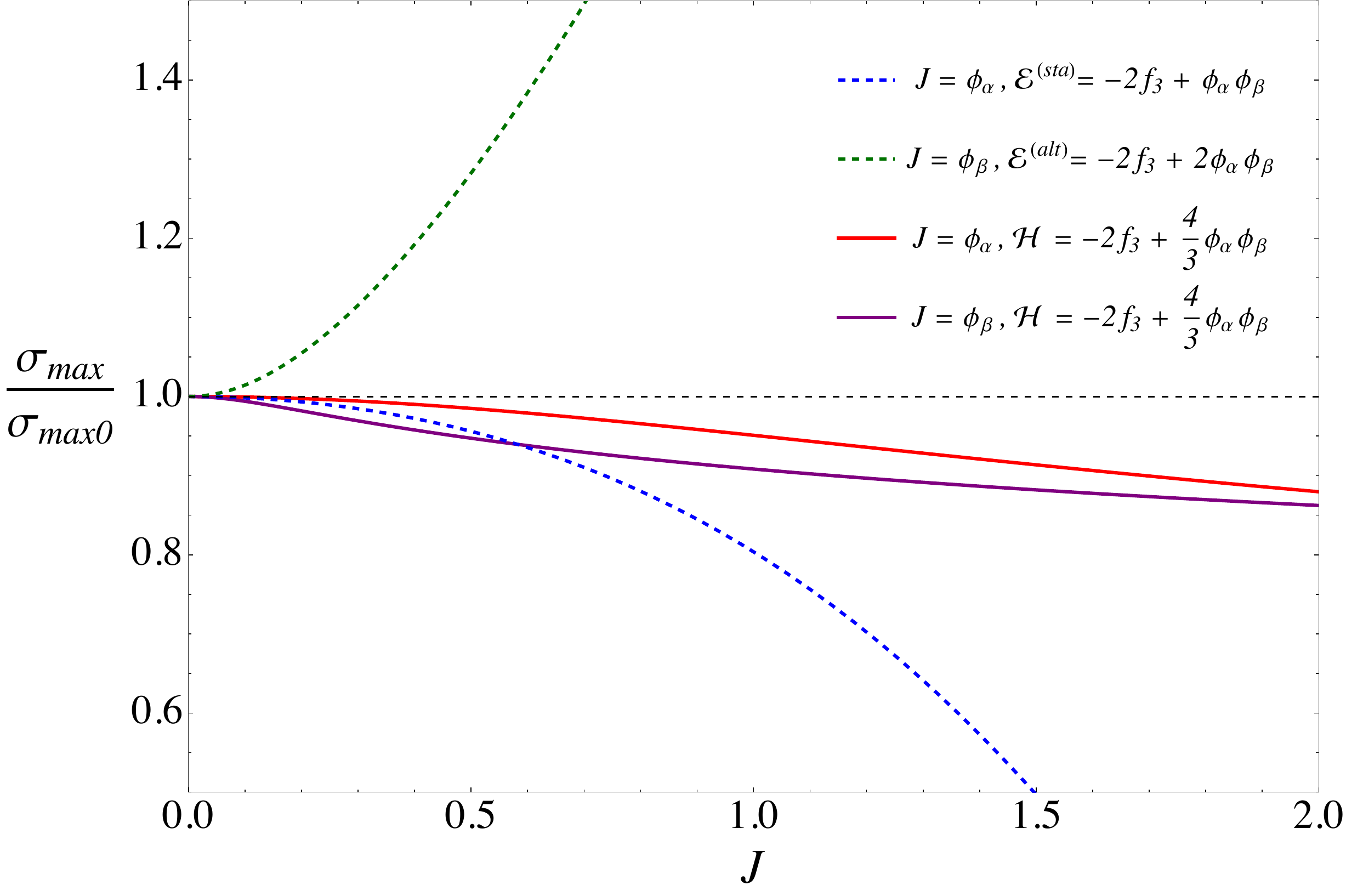}
	\caption{Late-time growth rate of holographic complexity $\sigma_{\text{max}}$ under the CV conjecture as a function of the source $J$ with fixed energies under different quantization schemes. The vertical axis shows the normalized growth rate $\sigma_{\text{max}} / \sigma_{\text{max}0}$, where $\sigma_{\text{max}0}$ corresponds to the vacuum black hole. The blue and green dashed curves correspond to the standard and alternative quantization schemes with fixed $\mathcal{E}^{\text{(sta)}}$ and $\mathcal{E}^{\text{(alt)}}$, respectively. The red and purple solid curves represent cases with fixed modified energy $\mathcal{H}$.}
	\label{Fig4002}
\end{figure}
As shown in Fig.~\ref{Fig4002}, the vertical axis $\sigma_{\text{max}} / \sigma_{\text{max}0}$ represents the normalized growth rate, where $\sigma_{\text{max}0}$ corresponds to the vacuum black hole (i.e., with vanishing scalar field). When the modified energy $\mathcal{H}$ is held fixed, $\sigma_{\text{max}}$ is maximized precisely when the scalar field vanishes. This demonstrates that the Lloyd's bound for complexity growth is restored when the total energy is defined by the modified energy $\mathcal{H}$.

\subsection{Growth Rate of Complexity in CA Conjecture}\label{sec:CA_complexity}
In addition to the Complexity-Volume (CV) conjecture, holographic complexity also includes the Complexity-Action (CA) conjecture~\cite{Brown:2015bva,Brown:2015lvg}. In this section, we investigate the growth rate of holographic complexity under the CA framework and examine its maximal value for a fixed total energy. According to the CA conjecture, the complexity of a boundary field theory is proportional to the action of the Wheeler-DeWitt (WdW) patch in the bulk spacetime:
\begin{equation}
	\label{eq:400041}
	\mathcal{C}_{A} = \frac{I_{\text{WdW}}}{\pi \hbar}.
\end{equation}
The Wheeler-DeWitt (WdW) action $I_{\text{WdW}}$ consists of the following contributions:
\begin{equation}
	\label{eq:400042}
	I_{\text{WdW}} = I_{\text{bulk}} + I_{\text{bdy}} + I_{\text{joint}}.
\end{equation}
Here, $I_{\text{bulk}}$ represents the bulk action within the WdW patch, while $I_{\text{bdy}}$ accounts for contributions from the patch boundaries, including timelike, spacelike, and null surfaces, as well as the corresponding reparameterization-independent counterterms on null surfaces. $I_{\text{joint}}$ captures joint contributions arising from intersections between different boundary segments, such as junctions between two null boundaries or between a spacelike boundary (e.g., the singularity) and a null boundary.

To obtain a finite expression for the complexity, the action is regularized by introducing counterterms at the left and right vertices of the WdW patch~\cite{Yang:2017amx}. These counterterms are time-independent and therefore do not affect the complexity growth rate; accordingly, they will not be discussed in detail in this work.

One key evidence for the growth rate in the CA conjecture is that it satisfies the Lloyd's bound~\cite{Lloyd:2000cry}:
\begin{equation}
	\label{eq:400043}
	\frac{\text{d} \mathcal{C}}{\text{d} t} \leq \frac{2M}{\pi} \qquad \text{or} \qquad \frac{\text{d} I}{\text{d} t} \leq 2M.
\end{equation}
Similar to the CV conjecture, the Lloyd's bound is one necessary condition for the acceptable proposal of holographic complexity. This bound was first tested in the context of the CA conjecture by Reference~\cite{Brown:2015bva}, which examined the Lloyd bound for a Schwarzschild-AdS black hole, confirming that the bound is satisfied in this scenario. Subsequently, Refs.~\cite{Brown:2015bva,Brown:2015lvg,Lehner:2016vdi,Cai:2016xho} computed the complexity growth rate for charged AdS black holes, and their results consistently indicate that the Lloyd's bound continues to hold.

Reference~\cite{Yang:2016awy} first considered the inequality~\eqref{eq:400043} without assuming special symmetries and provided a rigorous proof showing that the bound holds when matter fields exist only outside the Killing horizon and satisfy the strong energy condition. It remains an open question whether inequality~\eqref{eq:400043} holds in more general situations where matter fields extend into the horizon, particularly, for the situation that two different quantization schemes can be used.

\begin{figure}[h]
	\centering
	\includegraphics[width=0.45\textwidth]{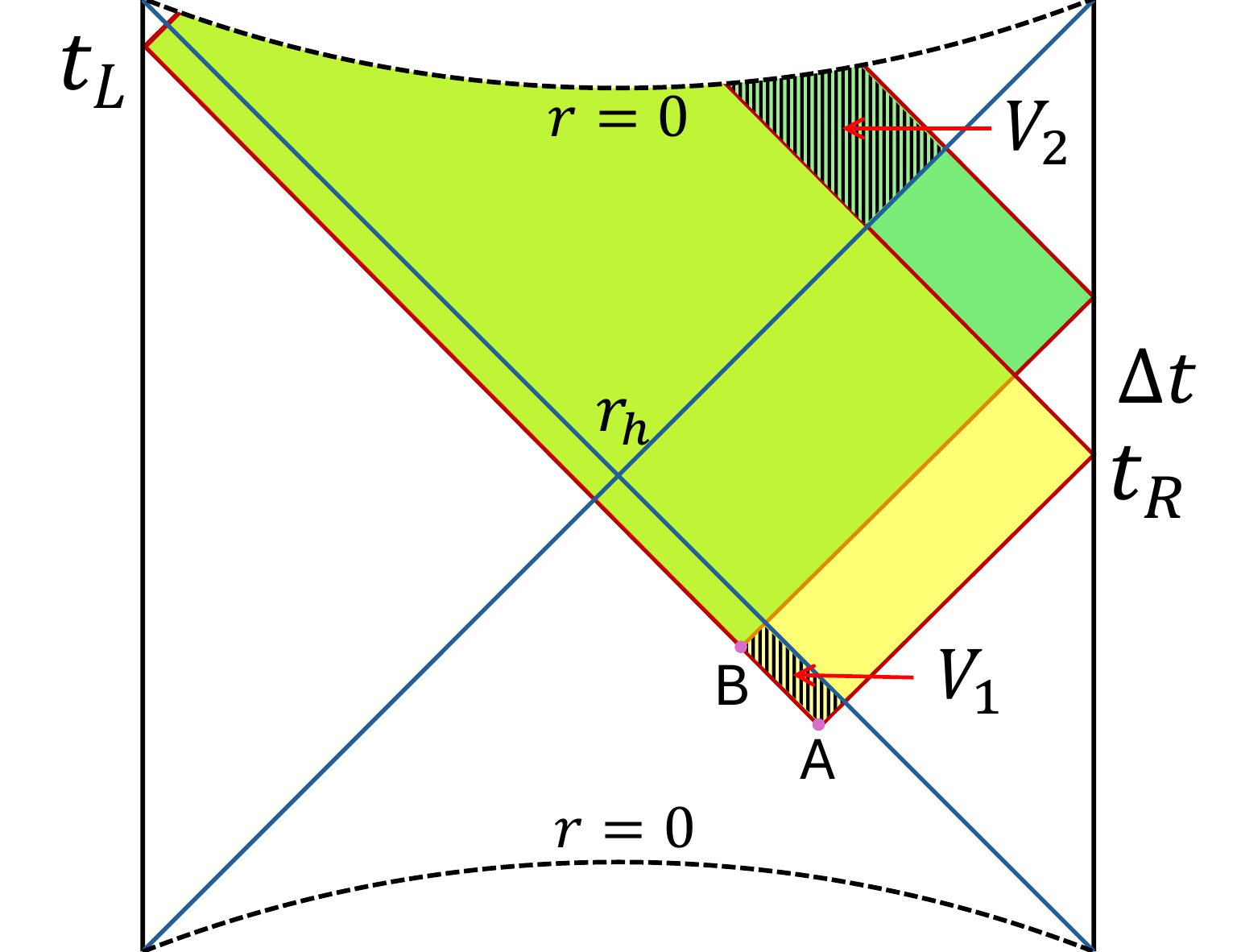}
	\caption{Wheeler-DeWitt patch at late time.}
	\label{Fig40042}
\end{figure}

We now provide the formula for computing the action growth rate in the late-time limit $t_{L} \rightarrow -\infty$, as illustrated in Figure~\ref{Fig40042}. We fix $t_L$ to be sufficiently large but finite, and shift $t_R \to t_R + \Delta t$. The variation of the on-shell action is then given by
\begin{equation}
	\label{eq:400044}
	\Delta I := I(t_L, t_R+\Delta t) - I(t_L, t_R) = \Delta I_{\mathrm{bulk}} + \Delta I_{\mathrm{bdy}} + \Delta I_{\mathrm{joint}}.
\end{equation}
Next, we compute the three contributions in Eq.~\eqref{eq:400044}. The difference between the two WdW patches corresponds to the yellow and green regions in Figure~\ref{Fig40042}. Due to symmetry, contributions from regions outside the horizon cancel, leaving only the subregions $V_1$ and $V_2$ to contribute to the action growth. In the late-time limit, the volume of $V_1$ is exponentially suppressed and can be neglected. Using the bulk metric~\eqref{eq:3002}, we obtain
\begin{equation}
	\label{eq:400045}
	\begin{aligned}
		\Delta I_{\mathrm{bulk}} &= -\frac{\Sigma_{d-1}}{8\pi} \oint_{\partial V_2} \mathrm{e}^{-\chi/2} f(r) r^{d-2} \, \mathrm{d}t
		= \frac{\Sigma_{d-1}}{8\pi} \lim_{r \to 0} \mathrm{e}^{-\chi/2} f(r) r^{d-2} \Delta t
		= -\frac{\Sigma_{d-1}}{8\pi} f_s \mathrm{e}^{-\chi_s/2} \Delta t, \\
		\Delta I_{\mathrm{bdy}} &= \Delta I_{\mathrm{GHY, singularity}}
		= \frac{\Sigma_{d-1} f_s \mathrm{e}^{-\chi_s/2}}{16\pi} \left( \frac{\phi_s^2}{d-2} + d \right) \Delta t, \\
		\Delta I_{\mathrm{joint}} &= I_{\mathrm{joint}}(B) - I_{\mathrm{joint}}(A)
		= \frac{\Delta t}{8\pi} \oint_{\mathcal{N}} \kappa \, \mathrm{d}S
		= T \mathcal{S} \, \Delta t.
	\end{aligned}
\end{equation}
Here, $f_s$, $\chi_s$, and $\phi_s$ are coefficients determined by the asymptotic behavior of the metric and scalar field near the spacetime singularity ($r \to 0$)~\cite{Cai:2020wrp}:
\begin{equation}
	\label{eq:4004}
	\begin{aligned}
		\phi(r) &= \phi_s \ln r + \cdots, \\
		f(r) &= -f_s r^{-\left[ d-2 + \frac{\phi_s^2}{2(d-1)} \right]} + \cdots, \\
		\chi(r) &= \chi_s - \frac{\phi_s^2}{d-1} \ln r + \cdots.
	\end{aligned}
\end{equation}
In Eq.~\eqref{eq:400045}, $\Delta I_{\mathrm{bulk}}$ receives contributions only from $V_2$. By applying Green's theorem, the bulk integral is transformed into a boundary integral over $\partial V_2$. The term $\Delta I_{\mathrm{bdy}}$ includes contributions from all boundaries; since null boundaries contribute trivially, only the Gibbons-Hawking-York (GHY) term on the spacelike boundary (the singularity) is relevant. Meanwhile, $\Delta I_{\mathrm{joint}}$ can be computed via the surface gravity at the bifurcated Killing horizon $\mathcal{N}$, as discussed in Ref.~\cite{Yang:2016awy}. Therefore, the late-time CA complexity growth rate can be expressed as
\begin{equation}
	\label{eq:400048}
	\lim_{t_L \to \infty} \frac{\mathrm{d} \mathcal{C}_A}{\mathrm{d} t_R} = \frac{\Sigma_{d-1}}{\pi} \left[ \frac{f_s \mathrm{e}^{-\chi_s/2}}{16 \pi} \left( \frac{\phi_s^2}{d-2} + d - 2 \right) + T \mathcal{S} \right].
\end{equation}

In the following, we investigate the relationship between the total energy of the spacetime and the growth rate of holographic complexity in the presence of a real scalar field. Notably, the conventional definition of total energy generally depends on the holographic renormalization scheme, implying that the validity of the bound may be scheme-dependent. In contrast, our modified definition of the total energy, $\mathcal{H}$, is scheme-independent, providing a new perspective on the interpretation and fulfillment of the bound. We again work within the model defined in Sec.~\ref{sec:example}. To study the dependence of the CA complexity growth rate on the scalar hair, we compute $\dot{\mathcal{C}}_A$ for configurations with fixed values of the standard energy $\mathcal{E}^{\text{(sta)}}$, the alternative energy $\mathcal{E}^{\text{(alt)}}$, and the modified energy $\mathcal{H}$.

\begin{figure}[h]
	\centering
	\includegraphics[width=0.7\textwidth]{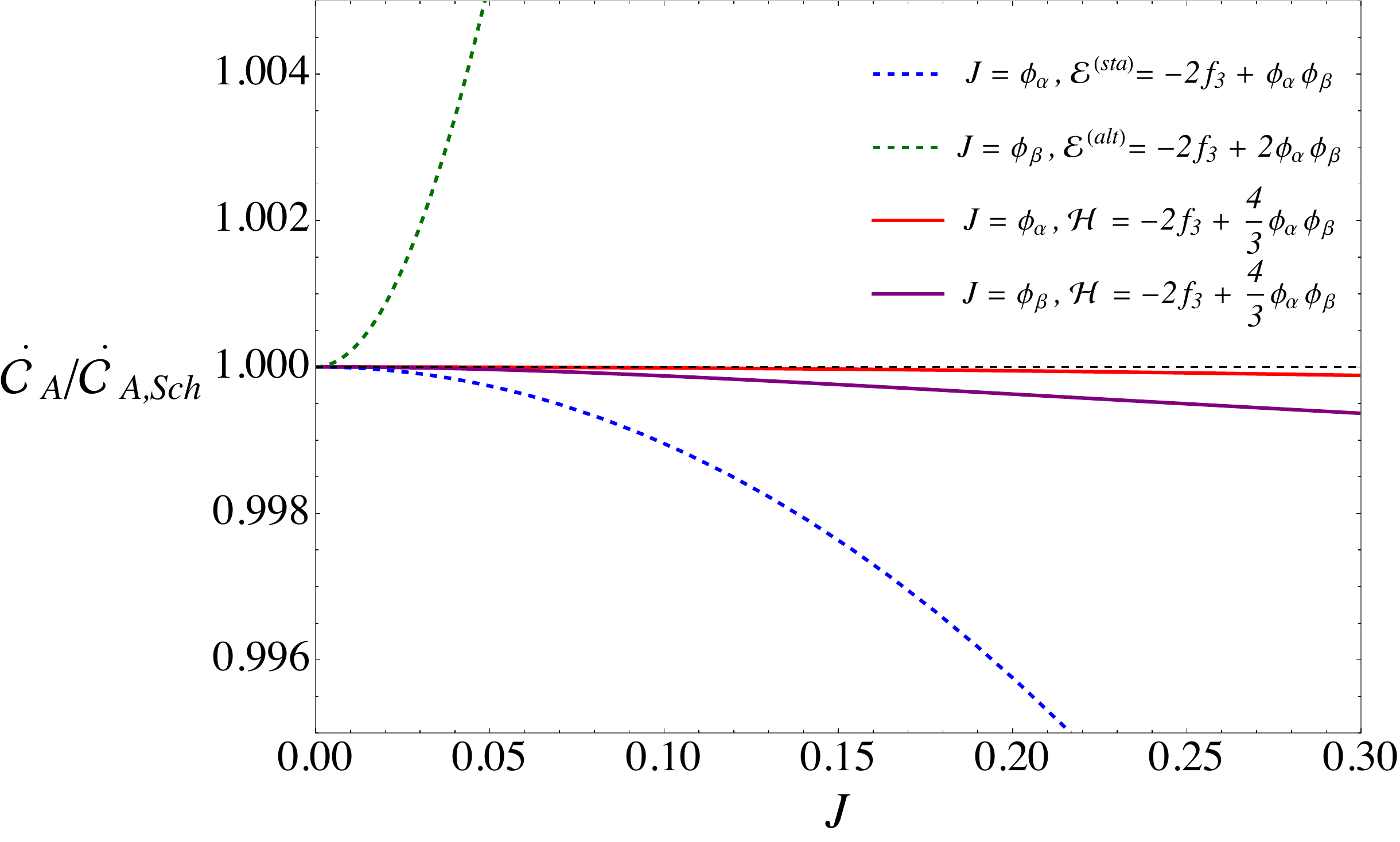}
	\caption{Late-time growth rate of holographic complexity $\dot{\mathcal{C}}_A$ under the CA conjecture as a function of the source  $J$ with fixed energies under different quantization schemes. The vertical axis shows the normalized growth rate $\dot{\mathcal{C}}_A / \dot{\mathcal{C}}_{A,\rm Sch}$, where $\dot{\mathcal{C}}_{A,\rm Sch}$ is the value for the vacuum black hole. The blue dashed and green dashed curves correspond to the standard and alternative quantization schemes with fixed $\mathcal{E}^{\text{(sta)}}$ and $\mathcal{E}^{\text{(alt)}}$, respectively. The red and purple solid curves represent cases with fixed modified energy $\mathcal{H}$.}
	\label{Fig4003}
\end{figure}

Figure~\ref{Fig4003} illustrates a key distinction. When the standard energy $\mathcal{E}^{\text{(sta)}}$ is held fixed, the vacuum black hole (with vanishing scalar field) yields the maximum late-time complexity growth rate, serving as an upper bound. In contrast, for a fixed alternative energy $\mathcal{E}^{\text{(alt)}}$, the vacuum configuration produces the minimum growth rate. Remarkably, when the modified energy $\mathcal{H}$ is held fixed, the vacuum black hole configuration again provides the maximum complexity growth rate $\dot{\mathcal{C}}_A$. This holds for the families of solutions corresponding to both quantization schemes (red and purple curves), demonstrating that $\mathcal{H}$ universally defines an energy for which the vacuum black hole has fastest complexity growth.

\section{Conclusion}\label{section 8}

In this paper, we have systematically examined the definition of total energy in AdS spacetime and its physical implications within the holographic duality framework. The definition of total energy in AdS has long posed theoretical challenges. In certain cases, the ADM mass can be employed, but it is not applicable in most scenarios. The advent of the AdS/CFT correspondence, however, has enabled a holographic approach to this problem, formalized in the framework of holographic renormalization. In particular, the $T_{tt}$ component of the holographic stress-energy tensor, computed via holographic renormalization, can be interpreted as the total energy of the bulk spacetime. This approach has proven highly effective, allowing us to compute the total energy for AdS black hole spacetimes with scalar hair.

However, when the scalar field mass $m$ lies within the range $m^2_{\text{BF}} \leq m^2 < m^2_{\text{BF}} + 1/\ell_{\text{AdS}}^2$, holographic renormalization allows two distinct quantization schemes, which in turn yield different values for the total energy. Though it is not surprising that same bulk geometry can correspond to two different energies in the viewpoint of holographic renormalization, this situation still presents significant challenges in several contexts, including the study of the Penrose inequality, the upper bound on the growth rate of holographic entanglement entropy, and the upper bounds on the growth rate of holographic complexity (both CV and CA conjectures). Depending on the chosen quantization scheme, one may reach opposite conclusions regarding these bounds.

To resolve this ambiguity, we introduce in this work a new definition of total energy, denoted by $\mathcal{H}$, which is independent of the choice of quantization scheme. By construction, $\mathcal{H}$ assumes the same value under both standard and alternative quantization. Using this modified energy, we re-examine the Penrose inequality, the upper bounds on the growth rate of holographic entanglement entropy, and the upper bounds on holographic complexity growth (for both the CV and CA conjectures). We find that, in all these cases, the corresponding inequalities and bounds are consistently satisfied regardless of the quantization scheme. These results demonstrate both the reasonableness and the universality of our definition of $\mathcal{H}$, and indicate that it provides a more robust, scheme-independent measure for addressing energy-related questions in holographic setups with scalar fields. The results in this paper imply that our modified definition of energy could be regarded as the intrinsic energy of bulk spacetime and may be favored in the studies of various holographic bounds involving the total energy/mass. We emphasize that our result should not be interpreted as establishing the universal validity of bound in arbitrary holographic systems. The status of complexity bounds is known to be subtle in charged, rotating, and higher-derivative gravitational theories, where the relevant thermodynamic quantity may differ from the total energy. Our result only shows that quantization-scheme dependence introduces an additional ambiguity that should be removed before discussing such bounds.

Looking ahead, we plan to further test the applicability of $\mathcal{H}$ in other contexts where the definition of total energy is crucial, such as in dynamical spacetimes or in the presence of more complex matter fields. We anticipate that this approach will contribute to a deeper understanding of energy definitions in AdS/CFT, and help clarify the significance of quantization-scheme independence in holographic physical bounds.

Finally, let us comment on the scope of our analysis. Throughout this work, we have focused on the single-trace deformation theory and so we only have two canonical quantization schemes for scalar fields within the Breitenlohner-Freedman window, namely the standard (Dirichlet) and alternative (Neumann) quantizations. These two cases correspond to well-defined conformal fixed points, for which the identifications of the source $J$, expectation value $\langle O\rangle$, and operator dimension $\Delta$ are unambiguous. For more general mixed boundary conditions, which are typically associated with multi-trace deformations in the dual field theory, the situation becomes more subtle. Such boundary conditions generally describe renormalization group flows between the two fixed points rather than a fixed conformal theory. As a result, the identifications of the source, operator expectation value, and scaling dimension may become scheme-dependent and scale-dependent away from the fixed points. For this reason, we have restricted our analysis to the two fixed-point quantizations where the ambiguity can be isolated most clearly. It would be interesting to investigate whether a generalized version of the quantization-scheme-independent energy proposed in this work can be extended tomulti-trace deformation theories. We leave this question for future study.

\acknowledgments
This work is supported by the Natural Science Foundation of China under Grant No. 12375051, No.~12575061; Tianjin University Self-Innovation Fund Extreme Basic Research Project Grant No. 2025XJ22-0014 and 2025XJ21-0007; Tianjin University Graduate Liberal Arts and Sciences Innovation Award Program (2023) No. B1-2023-005.

\appendix

\section{Bulk Conserved Charge as the Origin of $\mathcal{H}$}\label{appendix A}
We introduced a quantization-scheme-independent energy $\mathcal{H}$. Its physical origin is somewhat ambiguous, because a ``total energy'' concept should arise from global symmetries or conservation laws. Therefore, whether there exists a deterministic global derivation that naturally produces the same boundary value as $\mathcal{H}$ is a question worth investigating.
Ref.~\cite{Cai:2021obq} studied charged vector hairy black holes and constructed a Noether charge by using the scaling symmetry of the radially reduced action. Inspired by this method, we apply the same technique to the present system that involves a scalar field. We identify a scaling symmetry and derive a conserved Noether charge, denoted $Q(r)$. Surprisingly, the calculation shows that $Q(\infty)$ exactly equals the $\mathcal{H}$ constructed in the previous section.

We consider Einstein gravity minimally coupled to a real scalar field in asymptotically AdS spacetime,
\begin{equation}
	\label{A1001}
	S=\frac{1}{16\pi G}\int \mathrm{d}^{d+1} x \sqrt{-g}\left(R-2\Lambda-\frac{1}{2} \nabla^{\mu} \phi \nabla_{\mu} \phi-\frac{1}{2}m^2\phi^2\right),
\end{equation}
where $\Lambda=-\frac{d(d-1)}{2\ell_{\text{AdS}}^{2}}$ denotes the cosmological constant, and $m$ is the mass of the scalar field $\phi$. For convenience we introduce $z=1/r$ and rewritten the metric~\eqref{eq:3002} into
\begin{equation}
	\td s^2 = \frac{1}{z^2}\left[ -\tilde{f}(z)e^{-\chi(z)}\td t^2 + \frac{\td z^2}{\tilde{f}(z)} + \td \mathbf{x}_{d-1}^2 \right],
	\label{eq:metric_z}
\end{equation}
where $\tilde{f}=r^{-2}f=z^2f$ and we have set the AdS radius \(\ell_{\text{AdS}} = 1\) for simplicity.  The scalar field \(\phi\) is now a function of \(z\) only, \(\phi = \phi(z)\).

We substitute ansatz~\eqref{eq:metric_z} into the action \eqref{A1001} and obtain $S$ as a functional of $\{\tilde{f},\chi,\phi,\, \tilde{f}',\chi',\phi',\, \tilde{f}'',\chi'',\, z\}$.
\begin{equation}\label{actioneff1}
S_{\text{eff}}=\frac{\mathcal{V}_{d}}{16\pi G}\int\mathcal{L}_{\text{eff}}(\tilde{f},\chi,\phi,\, \tilde{f}',\chi',\phi',\, \tilde{f}'',\chi''; z) \mathrm{d}z \,,
\end{equation}
where the effective Lagrangian reads
\begin{equation}\label{actioneff2}
\begin{split}
\mathcal{L}_{\text{eff}}&=-\frac{ e^{-\chi/2}}{z^{d+1}}\left[d(d+1)  \tilde{f}+2 \Lambda +\frac12m^2\phi^2+z^2\tilde{f}\phi'^2\right]\\
&\quad+\frac{d e^{-\chi/2}}{ z^d }\left(2\tilde{f}'-\tilde{f}\chi'\right)+\frac{\tilde{f} e^{-\chi/2}}{z^{d-1}}\left(\frac{3\chi'}{2}\frac{\tilde{f}'}{\tilde{f}}-\frac{\chi'^2}{2}-\frac{\tilde{f}''}{\tilde{f}}+\chi''\right)\,,\\
\end{split}
\end{equation}
with $\mathcal{V}_{d}=\int \td t\td^{d-1}x$. It is clear that $\mathcal{L}_{\text{eff}}$ depends on five real functions $\{\tilde{f},\chi,\phi\}$, of variable $z$, and on the derivatives of $\{\tilde{f},\chi,\phi\}$ up to second order. Moreover, $\mathcal{L}_{\text{eff}}$ depends explicitly on $z$. Remarkably, $S$ is invariant under the following transformation
\begin{equation}
   \label{A0003}
	\begin{aligned}
		z &\;\to\; \lambda z, \\
		\chi(z) &\;\to\; \chi(z) - 2d\ln\lambda,
	\end{aligned}
\end{equation}
$\tilde{f}(z)$ and $\phi(z)$ remains unchanged.
By Noether's theorem, we obtain the corresponding conserved charge
\begin{equation}
	\mathcal{Q}_{\text{Noether}}(z) = \frac{(d-1)e^{\chi(z)/2}}{z^{d-1}}\, \bigl( \tilde{f}(z) e^{-\chi(z)} \bigr)' .
	\label{eq:Q_Noether_z}
\end{equation}
After transforming to the $r$ coordinate and multiplying by a constant, this charge becomes
\begin{equation}
	\mathcal{Q}_{\text{Noether}}=Q(r) := \frac{(d-1)r^{d+1}e^{\chi/2}}{d} \left( \frac{f e^{-\chi}}{r^2} \right)' .
	\label{eq:Q_def_r}
\end{equation}

Taking the example of $d = 3$, $m^2 = -2$ and using the expansion~\eqref{eq:3008}, we find that the boundary value of $Q(r)$ is
\begin{equation}
	\label{eq:3016}
	\lim_{r\to\infty} Q(r) = -2 f_3 + \frac{4}{3} \phi_{\alpha} \phi_{\beta},
\end{equation}
which is exactly equals $\mathcal{H}$. Thus we can tentatively conclude that we have found a bulk counterpart of the boundary $\mathcal{H}$, namely $Q(r)$, and that $Q(r)$ is a Noether conserved charge.

Note that the transformation~\eqref{A0003} reduces on the boundary metric, where only the $\td t^2$ component changes
\begin{equation}
   \label{A0007}
		\frac{-f(z) e^{-\chi(z)}}{z^2} \td t^2 \to \frac{-f(z) e^{-\chi(z)}}{z^2} \lambda^{2d-2} \td t^2  ,
\end{equation}
which is equivalent to a mere scaling transformation of time
\begin{equation}
   \label{A0008}
		t \to \lambda^{d-1} t
\end{equation}
on the dual boundary theory. Therefore, we may tentatively consider $\mathcal{H}$ as the conserved charge associated with time scaling $t \to \lambda^{d-1} t$ in the boundary field theory, even though our derivation is based on the gravity side rather than a direct field theory computation. Nevertheless, the bulk-to-boundary matching $Q(\infty)=\mathcal{H}$ strongly supports this interpretation, and it aligns with the general expectation that scaling symmetries in the bulk give rise to conserved charges dual to boundary scale transformations. It need to note the scaling transformation~\eqref{A0008} is different from the trivial scaling  transformation $(t,z,\vec{x})\rightarrow (\lambda t,\lambda z,\lambda \vec{x})$. We here focus on scalar theory. It is interesting to investigate if such coincident have deeper physical origin for other cases and we hope we could address this issue in the future.

\bibliographystyle{JHEP}

\bibliography{ref-ModifiedEnergy-1}

\end{document}